\documentclass[aps,prx,floats,floatfix,twocolumn,superscriptaddress,longbibliography]{revtex4-1}

\usepackage[utf8]{inputenc}
\usepackage{latexsym,amsmath}
\usepackage{amsbsy}
\usepackage[pdftex]{graphicx}
\usepackage{amssymb}
\usepackage{epstopdf}
\usepackage[usenames]{color}
\usepackage{float}
\usepackage{graphicx}
\usepackage{amsmath}
\usepackage[normalem]{ulem}
\usepackage{physics}
\usepackage{multibib}
\usepackage[colorlinks=true,%
bookmarks=false,%
linkcolor=blue,%
urlcolor=blue,%
citecolor=blue,%
breaklinks]{hyperref}

\begin{document}

\title{Learning to measure: adaptive informationally complete\\generalized measurements for quantum algorithms}

\author{Guillermo Garc\'{i}a-P\'{e}rez}
\affiliation{QTF Centre of Excellence, Department of Physics, Faculty of Science, University of Helsinki, Finland}
\affiliation{Complex Systems Research Group, Department of Mathematics and Statistics,
University of Turku, FI-20014 Turun Yliopisto, Finland}
\affiliation{Algorithmiq Ltd, Linnankatu 55 K 329, 20100 Turku, Finland}
\affiliation{InstituteQ - the Finnish Quantum Institute, University of Helsinki, Finland}

\author{Matteo A. C. Rossi}
\affiliation{Algorithmiq Ltd, Linnankatu 55 K 329, 20100 Turku, Finland}
\affiliation{QTF Centre of Excellence, Center for Quantum Engineering, Department of Applied Physics, Aalto University School of Science, Aalto FI-00076, Finland}
\affiliation{InstituteQ - the Finnish Quantum Institute, Aalto University, Finland}

\author{Boris Sokolov}
\affiliation{QTF Centre of Excellence, Department of Physics, Faculty of Science, University of Helsinki, Finland}
\affiliation{Algorithmiq Ltd, Linnankatu 55 K 329, 20100 Turku, Finland}
\affiliation{InstituteQ - the Finnish Quantum Institute, University of Helsinki, Finland}

\author{Francesco Tacchino}
\affiliation{IBM Quantum, IBM Research – Zurich, 8803 R\"{u}schlikon, Switzerland}

\author{Panagiotis Kl. Barkoutsos}
\affiliation{IBM Quantum, IBM Research – Zurich, 8803 R\"{u}schlikon, Switzerland}

\author{Guglielmo Mazzola}
\affiliation{IBM Quantum, IBM Research – Zurich, 8803 R\"{u}schlikon, Switzerland}

\author{Ivano Tavernelli}
\affiliation{IBM Quantum, IBM Research – Zurich, 8803 R\"{u}schlikon, Switzerland}

\author{Sabrina Maniscalco}
\affiliation{QTF Centre of Excellence, Department of Physics, Faculty of Science, University of Helsinki, Finland}
\affiliation{Algorithmiq Ltd, Linnankatu 55 K 329, 20100 Turku, Finland}
\affiliation{InstituteQ - the Finnish Quantum Institute, University of Helsinki, Finland}
\affiliation{QTF Centre of Excellence, Center for Quantum Engineering, Department of Applied Physics, Aalto University School of Science, Aalto FI-00076, Finland}
\affiliation{InstituteQ - the Finnish Quantum Institute, Aalto University, Finland}

\begin{abstract}
Many prominent quantum computing algorithms with applications in fields such as chemistry and materials science require a large number of measurements, which represents an important roadblock for future real-world use cases. We introduce a novel approach to tackle this problem through an adaptive measurement scheme. We present an algorithm that optimizes informationally complete positive operator-valued measurements (POVMs) on the fly in order to minimize the statistical fluctuations in the estimation of relevant cost functions. We show its advantage by improving the efficiency of the variational quantum eigensolver in calculating ground-state energies of molecular Hamiltonians with extensive numerical simulations. Our results indicate that the proposed method is competitive with state-of-the-art measurement-reduction approaches in terms of efficiency. In addition, the informational completeness of the approach offers a crucial advantage, as the measurement data can be reused to infer other quantities of interest. We demonstrate the feasibility of this prospect by reusing ground-state energy-estimation data to perform high-fidelity reduced state tomography.
\end{abstract}

\maketitle

\section{Introduction}

Quantum computing is a rapidly growing multidisciplinary field with a very clear objective: to understand if, and to what extent, it is possible to build computing machines able to perform tasks that are impossible for conventional (classical) computers. Theoretically, milestone discoveries such as Shor's and Grover's quantum algorithms hint toward a positive answer to this question. These algorithms, which exploit quantum properties of the processor, can in principle outperform all currently existing classical methods. In practice, however, the implementation of such protocols in the regimes of interest will most probably require the use of ideal fault-tolerant universal quantum computers. At the same time, because of the extreme fragility of quantum information storage and processing in the presence of environmental noise, error-correction techniques required to achieve fault tolerance are still experimentally in their infancy.

Universal fault-tolerant quantum computers, however, are not the only type of quantum machines able to tackle computationally hard problems. In fact, we can reformulate the main quantum computing research question and ask ourselves: what are the useful problems that quantum computers can solve more efficiently than their classical counterparts and, specifically, which subclasses of such problems are less demanding in terms of experimental requirements, given the current state-of-the-art quantum hardware? Note that this question has a different starting point, namely it focuses on our current ---or near-future--- technologies and devices, and aims at identifying, based on the current understanding, useful applications that may benefit from them.

There are at least two classes of problems that satisfy the above requirements. The first class has a long-standing history, dating back to Feynman (1982)~\cite{feynman_simulating_1982} and Manin (1980)~\cite{manin_computable_1980}, who pointed out that the simulation of quantum systems is hard on classical computers, while, under certain conditions, they can be efficiently investigated by means of other quantum systems~\cite{GeorgescuRMP2014}. In fact, this can be done using either digital quantum  simulators,  namely  specific-purpose  quantum computers~\cite{Lloyd1996,RevTacchino,NatPhysIBM},
or by employing analog quantum simulators~\cite{houck_-chip_2012,Monroe17,Lukin17,PRXQuantumSimulators2021}, namely other equivalent but easier-to-control quantum systems. %
The second class of problems emerges when we lift the requirement of finding ``exact'' solutions to a given problem.
Approximate near-term quantum devices might be able, e.g., to find better solutions to certain worst-case instances of non- deterministic polynomial-time hard (NP-hard) problems or find such approximate solutions faster.

A final ingredient to move toward the existing
approximate noisy quantum
devices~\cite{preskill2018quantum,bharti2021noisy} is the combination of quantum and classical techniques to maximize performance. In this paper, we focus on variational quantum algorithms, which have emerged recently as the most suited paradigm to tackle
the classes of problems identified above~\cite{Moll_2018,cerezo2020variational} with approximate quantum computing.
Specifically, these protocols are implemented by preparing a parametrized $N$-qubit trial state on a quantum device, extracting some observable quantities with suitable measurements and processing such measurement outcomes using a classical optimizer. The latter then returns the small changes that need to be implemented to prepare, in the next step, an updated trial wave function. This cycle is repeated many times until it converges to a quantum state from which the desired approximate solution can be extracted.

This procedure can be used to solve problems in chemistry~\cite{Kandala17,barkoutsos_particle_hole_2018,McArdleRMP2020,OllitraultPRR2020}, for the design of new materials~\cite{BarkoutsosAlchemical2021}, and generally in every field of physics where one needs to extract the properties of many-body quantum correlated systems, e.g., interacting fermionic systems, which are typically hard to simulate on classical devices~\cite{MazzolaPRL2019,CaiPRAppl2020}. In this case, these algorithms go by the name of Variational Quantum Eigensolvers (VQE)~\cite{PeruzzoVQE,McClean16,Kandala17}. In essence, the quantum processor is used to explore the exponentially large Hilbert space of the fermionic particles in order to find iteratively the ground state of the Hamiltonian, without solving the full diagonalisation problem. As an example, the knowledge of the ground state of a chemical compound as a function, e.g., of the nuclear coordinates allows one to extract crucial information such as the equilibrium bond length, bond angle, and dissociation energy. Note that, at least in principle, a quantum computer with a few hundreds of qubits could already have the potential to solve useful quantum chemistry problems that are intractable on classical computers.

The application of VQE has already been demonstrated in many proof-of-principle experiments~\cite{PeruzzoVQE,Kandala17,HempelPhysRevX2018,Kandala19,GanzhornPhysRevApplied2019}. However, a few major challenges still need to be overcome along the path to useful quantum advantage. On one hand, the classical optimization step that is associated with variational quantum algorithms can in general incur high computational costs because of the existence of many local minima or due to the problem of vanishing gradients~\cite{mcclean_barren_2018}. Some possible solutions have been proposed, combining techniques borrowed from classical optimization theory with a careful design of the variational ansatz, such as the recently proposed ADAPT-VQE~\cite{grimsley_adaptive_2019} and oo-VQE~\cite{Sokolov2020b}, and of the associated cost function~\cite{cerezo_cost_2021}. On the other hand, the so-called measurement problem arises from the very high cost in terms of the number of observations that are typically needed to reconstruct the properties of interest, and specifically, the expectation value of the Hamiltonian, on the quantum states constructed by variational means. In fact, as the size of the problem approaches the regime in which the VQE could compete with classical methods, the current approaches would lead to prohibitive requirements to reach the desired degree of accuracy~\cite{McCleanLocality2014,wecker2015progress,BabbushPhysRevX2018,CaiPRAppl2020}.

In this work, we tackle the second problem, by presenting a novel adaptive method that sensibly alleviates the demands on the number of measurements, thus paving the way for an increase of the affordable problem sizes in experimental realisations.

On a fundamental level, our approach introduces a new perspective on how to improve the overall observables reconstruction strategy in VQE, and possibly in variational algorithms in general, by leveraging informationally complete quantum measurements.

Before introducing our protocol, however, in the next section we describe the measurement problem in more detail and briefly mention the main approaches that have been proposed in the literature to tackle it in the next section.

\section{The measurement problem}\label{sec:measurement_problem}
One of the most prominent differences between classical and quantum methods concerns the way in which information is extracted at the end of the execution of the algorithm. In a typical situation, the quantum circuit prepares a $N$-qubit quantum state $\vert\psi \rangle$ that is used to compute the expectation value of an operator $\langle \mathcal{O} \rangle = \langle \psi \vert \mathcal{O} \vert \psi \rangle$. Generally, it is not possible to measure $\mathcal{O}$ directly in its eigenbasis. For instance if we are interested in finding the ground state of the Hamiltonian $H$, measuring in its eigenbasis requires solving the problem itself in advance.

The standard measurement protocol, henceforth named the \emph{Pauli} method, consists in writing the operator $\mathcal{O}$ as a sum of $K$ Pauli strings, $\mathcal{O} = \sum_\mathbf{k} c_\mathbf{k} P_\mathbf{k}$, where  $P_\mathbf{k} = \bigotimes_{i=1}^{N} \sigma_{k_i}^{(i)}$ and  $\sigma_0^{(i)} = \mathbb{I}^{(i)}, \sigma_1^{(i)} = \sigma_x^{(i)}, \dots$ are Pauli operators. The expectation value of the operator is therefore obtained in terms of the weighted sum of $K$ expectation values, $\langle \mathcal{O} \rangle = \sum_\mathbf{k} c_\mathbf{k} \langle P_\mathbf{k} \rangle$.

Unfortunately, this method leads to a suboptimal measurement scheme, as the variance of $\mathcal{O}$ is the sum of the weighted variances of the individual operators $P_\mathbf{k}$. More precisely, the error in the estimation is given by
\begin{equation}
 \epsilon=\sqrt{\sum\limits_\mathbf{k} \vert c_\mathbf{k} \vert^2 \mathrm{Var}(P_\mathbf{k})/S_\mathbf{k}},
\end{equation}
where $\mathrm{Var}(P_\mathbf{k}) = \langle P_\mathbf{k}^2 \rangle - \langle P_\mathbf{k} \rangle^2$ is the variance
of $P_\mathbf{k}$ and $S_\mathbf{k}$ is the number of measurements, i.e., wave-function collapses, used to estimate term $\mathbf{k}$~\cite{wecker2015progress}. Interestingly, under such measurement scheme, even exactly prepared ground states do not enjoy the \emph{zero-variance} property, such that statistical energy fluctuations always remain finite and large.

This constitutes a major source of problems for variational-based state preparation, where circuit parameters are optimized to minimize the expectation value of the energy. Given its significance, several efforts have been put forward to mitigate this problem. One simple strategy, henceforth named \emph{grouped Pauli} method, aims at identifying all the Pauli strings that can be measured simultaneously from the same data set~\cite{Kandala17}. While this is not solving the issue, it reduces the computational overhead of the procedure. Promising approaches also involve the usage of a classical machine-learning engine to perform an \emph{approximate} reconstruction of the quantum state~\cite{torlai_2018_nnqst} using only the basis state  defined by $P_\mathbf{k}$~\cite{torlai2020precise}, or \emph{classical shadows} of a quantum state~\cite{huang2020predicting,hadfield2020measurements,huang2021efficient}. Other approaches based on grouping of commuting terms, effective measurement scheduling and optimized qubit tomography have been described in Refs.~\cite{Jena2019,Yen2019b,Huggins2019,Gokhale2019,Crawford2019,zhao2019measurement,paini2019approximate,BonetPhysRevX2020,CotelrPRL2020,hamamura_efficient_2020}. In the context of quantum state tomography with generalized quantum measurements, neural network-assisted adaptive methods have also been proposed~\cite{quek_adaptive_2021}. In the following, we show how a related idea can be applied to fully general observable reconstruction tasks and gradient-based measurement learning with effective sampling costs.
It is also worth reminding that, in a more general scenario in which fault-tolerant architectures are available, optimal strategies for obtaining expectation values with
Heisenberg-limited precision are known, based on quantum phase estimation~\cite{Knill_measurements_2007}. Intermediate solutions between the standard and the quantum-phase-estimation-like sampling regimes are also possible, leveraging
some trade-offs between sample complexity and quantum coherence~\cite{Wang_alphaVQE_2019,Wang_prxquantum_2021}.

In this work, we present an algorithm for efficient observable estimation that exploits generalized quantum measurements integrating three important components: a hybrid quantum-classical Monte Carlo, a method to navigate generalized measurement space toward efficient measurements, and a recipe to combine different estimations of the observable of interest. The result is a procedure in which the optimal measurement of an operator average is learnt in an adaptive fashion with no measurement overhead.

\section{Adaptive measurement scheme}

In this section, we explain our adaptive measurement scheme. In a nutshell, the idea is to use parametric informationally complete positive operator-valued measures (IC POVMs), which can in principle be used to estimate any expectation value of our choice. We first introduce a hybrid Monte Carlo approach, which bypasses the need to use tomographic reconstructions of quantum states from the IC data. We then describe how, by using parametric families of POVMs, the measurement settings can be optimized to yield low statistical errors in the estimation of the target expectation values.

\begin{figure}[t]
    \centering
    \includegraphics[width=.95\columnwidth]{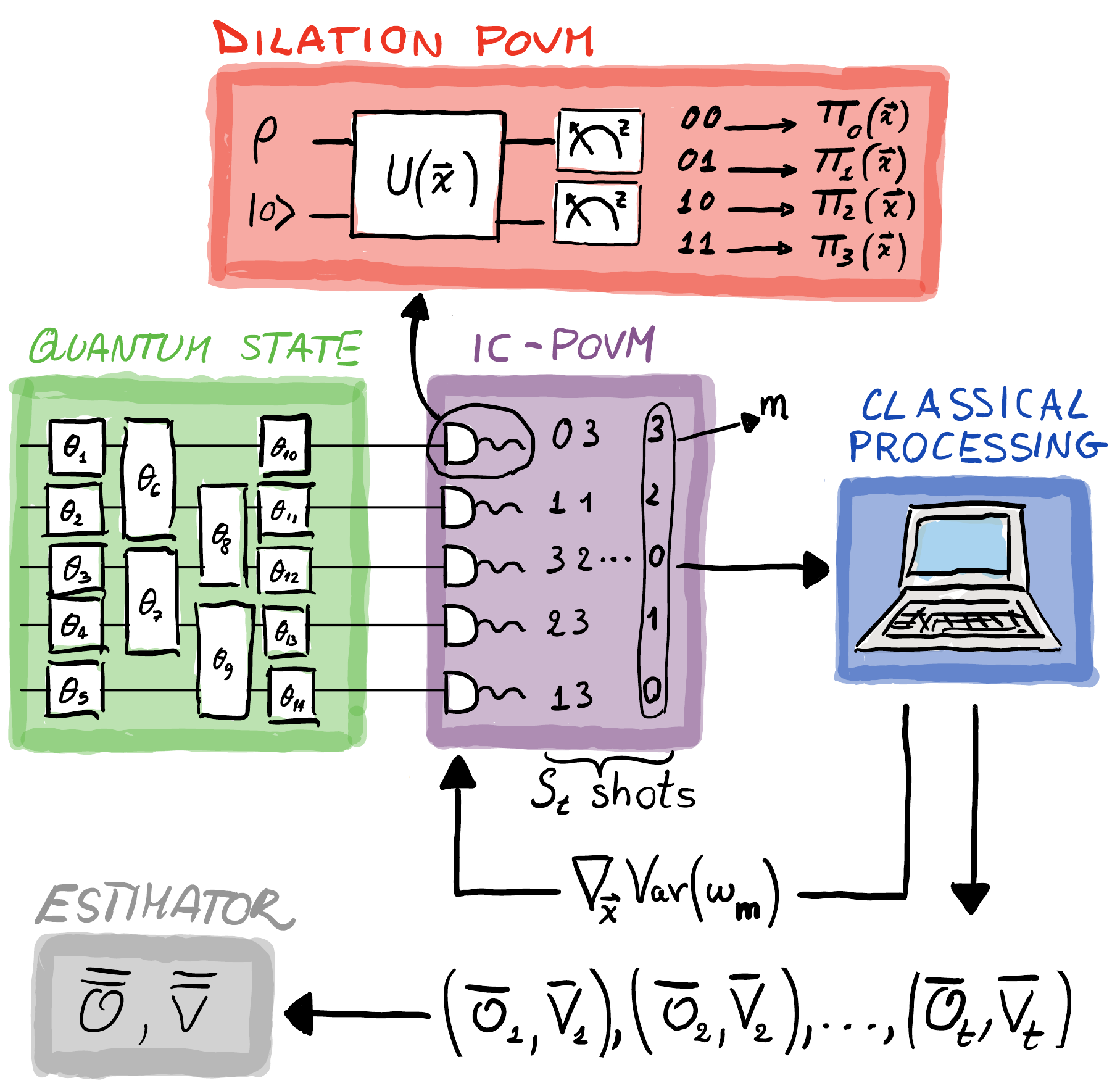}
    \caption{\textbf{Adaptive measurement scheme.} The sketch represents a typical measurement step of a variational algorithm. The ansatz prepares a state $| \psi (\vec{\theta}) \rangle$ (green box) for which the mean of some observable $\mathcal{O}$ must be evaluated. Our algorithm is an efficient measurement subroutine in this process. It relies on parametric informationally complete POVMs (purple box) implemented with ancillary qubits (red box). These are explained in detail in App.~\ref{app:povm_implementation}. Initially, we start by performing $S_1$ measurements using the POVM corresponding to parameters $\vec{x}_1$, and obtain $S_1$ outcomes $\mathbf{m}_1, \ldots, \mathbf{m}_{S_1}$. The measurement data are post-processed efficiently on a classical device (blue box) twice, with two different goals. First, we estimate the mean of the observable, $\bar{\mathcal{O}}_1$, and the corresponding error of the estimation, $\bar{V}_1$, as explained in Sec.~\ref{sec:monte_carlo}. Second, we calculate the gradient of the estimation variance, $\nabla_{\vec{x}} \textrm{Var} (\omega_\mathbf{m})$, in POVM parameter space, and thus find a better POVM for iteration 2 (see Sec.~\ref{sec:gradient_estimation} and App.~\ref{app:gradient_descent}). At every step $t$, the variables $\bar{\bar{\mathcal{O}}}$ and $\bar{\bar{V}}$ integrate all the estimations for $t' \leq t$ while minimising the overall statistical error (see Sec.~\ref{sec:on_the_fly} and App.~\ref{app:sequential_vs_one-step}). The process is repeated iteratively until $\bar{\bar{V}}$ is below some desired threshold.}
    \label{fig:sketch}
\end{figure}

With respect to the second point, special attention must be devoted to achieving the desired POVM optimization without incurring additional overheads in terms of, e.g., the number of repetitions (also named \textit{shots} in the following) of the state-preparation-and-measurement routine. As we will explain in the following, an adaptive method --- that is, an on-the-fly optimization --- will serve this scope. In brief, the key is to use the IC data obtained with one given POVM twice. First, we use them to produce an estimation of the mean of the observable. Second, the same set of results can also be employed to find a better POVM for the next experiment. The collection of intermediate estimators of the target observable, each constructed along the process with a different POVM, is finally integrated together as to minimize the overall statistical uncertainty. As a result of this strategy, the measurement learning procedure improves over the initial POVM (which turns out to be already quite efficient, as shown in Sec.~\ref{sec:results}) with no additional measurement costs. The scheme is illustrated and summarized in Fig.~\ref{fig:sketch}.

It is important to stress that the method does not require any approximations whatsoever. In fact, it is completely agnostic to the nature of the operator $\mathcal{O}$ to be measured, as long as it is given in terms of a linear combination of products of single-qubit observables (e.g., Pauli strings). While the algorithm is rather general, its performance is strongly dependent on the weight of such products (the number of non-identity single-qubit operators in every term), as we explain later, which makes quantum chemistry with low-weight fermion-to-qubit mappings, such as Bravyi-Kitaev~\cite{bk2002} and the one recently introduced in Ref.~\cite{Jiang2020optimalfermionto}, ideal use cases. Moreover, it should be mentioned that the methodology relies on the use of one ancillary qubit for every system qubit. However, the ancillary qubits remain in the ground state until the measurement stage, and the procedure only requires an increase in the circuit depth that is independent of the system size (i.e., a single layer of two-qubit gates that can all be executed in parallel). Yet, the efficiency of the method when applied to quantum chemistry problems is comparable to that of state-of-the-art methods that require an additional circuit depth linear in the number of qubits~\cite{Huggins2019} and, at the same time, it provides informationally complete (IC) data useful for purposes beyond energy estimation.

To ease the explanation of the algorithm, we present its three main components separately. We first introduce the hybrid quantum-classical Monte Carlo sampling for the estimation of expectation values of operators in Sec.~\ref{sec:monte_carlo}. We then show in Sec.~\ref{sec:gradient_estimation} how to estimate the gradient in the space of POVMs without additional measurements, using only efficient classical post-processing and, lastly, in Sec.~\ref{sec:on_the_fly}, we illustrate how to integrate all the data obtained from different POVMs to estimate mean values while minimising statistical fluctuations.

\subsection{Hybrid quantum-classical Monte Carlo sampling}\label{sec:monte_carlo}

Our proposed algorithm relies on single-qubit (minimal) IC POVMs, which can be realised by applying a two-qubit gate between a system qubit and an ancillary one, the latter in a known state, and then measuring both qubits in the computational basis. In practice, this means that the ancillary qubits are initialised along with all the other qubits in the device (e.g., they are prepared in the ground state $\ket{0}$) and no operations are applied to them until the measurement stage. The implementation of these POVMs on current quantum computers has recently been demonstrated experimentally on IBM Quantum devices~\cite{oszmaniec2019simulating,garcia2021experimentally}.

By definition, one such POVM is represented by four linearly independent positive operators $\lbrace \Pi_i > 0, \, i = 0, \ldots, 3 \rbrace$ adding up to identity, $\sum_i \Pi_i = \mathbb{I}$, and spanning the space of linear operators in the Hilbert space $\mathcal{H}$ of the system qubit. Each of these operators, usually called \textit{effects}, is associated with one of the four possible outcomes of the two-qubit measurement, with $\mathrm{Tr} [ \rho \Pi_i ]$ being the probability of outcome $i$ on the quantum state $\rho$ of the target qubit. It is important to note that different qubit-ancilla unitaries generally lead to different POVMs. Hence, by parametrising these unitaries, we can parametrize the corresponding family of POVMs (see App.~\ref{app:povm_implementation}).

Let us consider the $N$-qubit case, with local and not necessarily identical POVMs associated with each qubit. The four effects associated with qubit $i$ are denoted by $\Pi_m^{(i)}$, with $m$ running from 0 to 3. The outcome of an experiment in which all qubits are measured via these local POVMs is a string $\mathbf{m} = (m_1, \ldots, m_N)$, where $m_i \in \lbrace 0, \ldots, 3 \rbrace$. The probability of such outcome given an $N$-qubit state $\rho$ is $\mathrm{Tr [ \rho \Pi_{\mathbf{m}}}]$, with $\Pi_{\mathbf{m}} = \bigotimes_{i=1}^{N} \Pi_{m_i}^{(i)}$. The set of these $4^N$ effects $\Pi_{\mathbf{m}}$ is IC in $\mathcal{H}^{\otimes N}$.

As explained in previous sections, in VQE realisations one typically needs to measure an operator $\mathcal{O}$ that can be decomposed in terms of $K$ Pauli strings, $\mathcal{O} = \sum_\mathbf{k} c_\mathbf{k} P_\mathbf{k}$ (we assume $c_\mathbf{k} \in \mathbb{R}$, as is customary, although our results can be easily generalized to complex-valued coefficients). Given that each of the local POVMs is IC, we can express the Pauli operators acting on each qubit $i$ in terms of the effects $\Pi_m^{(i)}$ as $\sigma_{k}^{(i)} = \sum_m b_{km}^{(i)} \Pi_{m}^{(i)}$, with which we can write
\begin{equation}\label{eq:decomposition}
\begin{aligned}
\mathcal{O} &= \sum_\mathbf{k} c_\mathbf{k} \bigotimes\limits_{i=1}^{N} \sigma_{k_i}^{(i)}
= \sum_\mathbf{k} c_\mathbf{k} \bigotimes\limits_{i=1}^{N} \left(\sum\limits_{m_i = 0}^{3} b_{k_i m_i}^{(i)} \Pi_{m_i}^{(i)} \right) \\
&= \sum_\mathbf{m} \sum_\mathbf{k} c_\mathbf{k} \prod\limits_{i=1}^{N} b_{k_i m_i}^{(i)} \Pi_\mathbf{m} \equiv \sum_\mathbf{m} \omega_\mathbf{m} \Pi_\mathbf{m}.
\end{aligned}
\end{equation}

The above expression seems useless at first sight: we transform a representation of $\mathcal{O}$ in terms of $K$ terms $c_\mathbf{k} P_\mathbf{k}$ into one with possibly $4^N$ terms $\omega_\mathbf{m} \Pi_\mathbf{m}$. However, the expectation value of the operator now reads
\begin{equation}
\langle \mathcal{O} \rangle = \mathrm{Tr} [ \rho \mathcal{O} ] = \sum_\mathbf{m} \omega_\mathbf{m} \mathrm{Tr} [ \rho \Pi_\mathbf{m} ] = \sum_\mathbf{m} \omega_\mathbf{m} p_\mathbf{m},
\end{equation}
where $p_\mathbf{m}$ is the probability of obtaining outcome $\mathbf{m}$. In other words, the mean value of the operator is the average of $\omega_\mathbf{m}$ over the probability distribution $\lbrace p_\mathbf{m}\rbrace$, $\langle \mathcal{O} \rangle = \langle \omega_\mathbf{m} \rangle_{\lbrace p_\mathbf{m}\rbrace}$. This observation enables a very different strategy for estimating $\langle \mathcal{O} \rangle$ as compared to the standard Pauli method introduced in Sect.~\ref{sec:measurement_problem}. Instead of evaluating each of the $p_\mathbf{m} = \langle \Pi_\mathbf{m} \rangle$ via repeated sampling, and once \emph{all} these mean values are known with high enough precision calculating $\langle \mathcal{O} \rangle = \sum_\mathbf{m} \omega_\mathbf{m} \langle \Pi_\mathbf{m} \rangle$, which would be infeasible given the aforementioned exponential amount of terms, we can resort to a Monte Carlo approach.

In Monte Carlo integration, one can evaluate integrals over high-dimensional domains efficiently by randomly sampling points within the domain and averaging their image through a suitable function. Similarly, in our case, we can exploit the fact that $p_\mathbf{m}$ is the probability of the measurement yielding outcome $\mathbf{m}$ to calculate $\langle \mathcal{O} \rangle$ in a similar manner, that is, using the quantum computer to sample values of $\mathbf{m}$ and a classical one to calculate the corresponding $\omega_\mathbf{m}$, hence bypassing the need to evaluate the mean values $\langle \Pi_\mathbf{m} \rangle$.

More precisely, the strategy is to repeat the measurement $S$ times using the local POVMs to sample from the probability distribution $\lbrace p_\mathbf{m} \rbrace$, resulting in a sequence of outcomes $\mathbf{m}_1, \ldots, \mathbf{m}_S$, and compute
\begin{equation}\label{eq:estimated_mean}
\bar{\mathcal{O}} = \frac{1}{S} \sum\limits_{s=1}^{S} \omega_{\mathbf{m}_s}.
\end{equation}
Each term $\omega_{\mathbf{m}_s} = \sum_\mathbf{k} c_\mathbf{k} \prod_{i=1}^{N} b_{k_i m_i}^{(i)}$ can be calculated in a polynomial time on a classical computer. This estimator converges to $\langle \mathcal{O} \rangle = \sum_\mathbf{m} \omega_\mathbf{m} p_\mathbf{m}$ as $\sqrt{\mathrm{Var} (\omega_\mathbf{m}) / S}$, where $\mathrm{Var} (\omega_\mathbf{m})$ is the variance of $\omega_\mathbf{m}$ over the probability distribution $\lbrace p_\mathbf{m} \rbrace$, hence possibly providing accurate estimations even when the sum in Eq.~\eqref{eq:estimated_mean} only involves a number of terms $S \ll 4^N$. Crucially, this method estimates the weighted average of all the Pauli strings $P_\mathbf{k}$ simultaneously, regardless of whether they commute or not, by exploiting IC data, yet circumventing any costly tomographic reconstruction of quantum states. In addition, in this Monte Carlo approach, the variance naturally takes into account the covariance between all these parallel measurements. In other words, the quantity $\sqrt{(\langle \omega_\mathbf{m}^2 \rangle_{\lbrace p_\mathbf{m}\rbrace} - \langle \omega_\mathbf{m} \rangle^2_{\lbrace p_\mathbf{m}\rbrace}) / S}$, which can be estimated efficiently from the data, accounts for the total statistical error. As we explain next, our strategy is to iteratively search for POVMs that minimize this error.

Importantly, the previous result holds for any operator $\mathcal{O}$, that is, the same sequence of outcomes $\mathbf{m}_1, \ldots, \mathbf{m}_S$ can be used to estimate, using only classical post-processing, any expectation value. However, not all expectation values can be estimated with the same precision. In particular, note that the products $\prod_{i=1}^{N} b_{k_i m_i}^{(i)}$ can in principle result in variances scaling exponentially in $N$. This is so because, generally, each coefficient $b_{k_i m_i}^{(i)}$ can have an absolute value different from, and also larger than, one. Hence, in worst-case scenarios, the absolute value of such products, and therefore of the $\omega_\mathbf{m}$ defined in terms of linear combinations of them, can scale unfavorably with the system size (see App.~\ref{app:sic-povms} for a concrete example). This limitation can be overcome for fermionic problems by using fermion-to-qubit mappings such as the Bravyi-Kitaev (BK)~\cite{bk2002} and especially the one recently proposed by Jiang et al. in Ref.~\cite{Jiang2020optimalfermionto}  (to which we refer as JKMN mapping), which lead to Pauli strings with logarithmic weight (that is, such that fermionic creation/annihilation operators are mapped onto Pauli strings with at most a logarithmic number of non-identity Pauli operators). Since the terms $b_{0 m_i}^{(i)}$, corresponding to the decomposition of identity, are always equal to one (recall that $\sum_m \Pi_m^{(i)} = \mathbb{I}^{(i)}$), these mappings guarantee that the products $\prod_{i=1}^{N} b_{k_i m_i}^{(i)}$ scale polynomially in $N$.

In any case, it should be clarified that using other mappings does not necessarily imply an unfavorable scaling of the algorithm, as there may nevertheless exist POVM parameters for which the method is efficient. In fact, as we show in Sect.~\ref{sec:results}, the adaptive strategy that we present in what follows finds POVMs for which the algorithm outperforms the Pauli and grouped Pauli methods in evaluating the ground-state energy of molecular Hamiltonians using the parity~\cite{Bravyi17} and Jordan-Wigner (JW) mappings as well.

Regarding the method proposed in Ref.~\cite{Jiang2020optimalfermionto}, it should be mentioned that our Monte Carlo approach, given in Eq.~\eqref{eq:estimated_mean}, offers some advantages over the latter. On the one hand, it bypasses the classical overhead needed for tomographic reconstructions. On the other hand, and more importantly, our approach does not disregard the covariance-induced statistical errors in the estimation of the average resulting from parallel measurements. These points are discussed in more detail in App.~\ref{app:sic-povms}.

\begin{figure*}[t!]
    \centering
    \includegraphics[width=\textwidth]{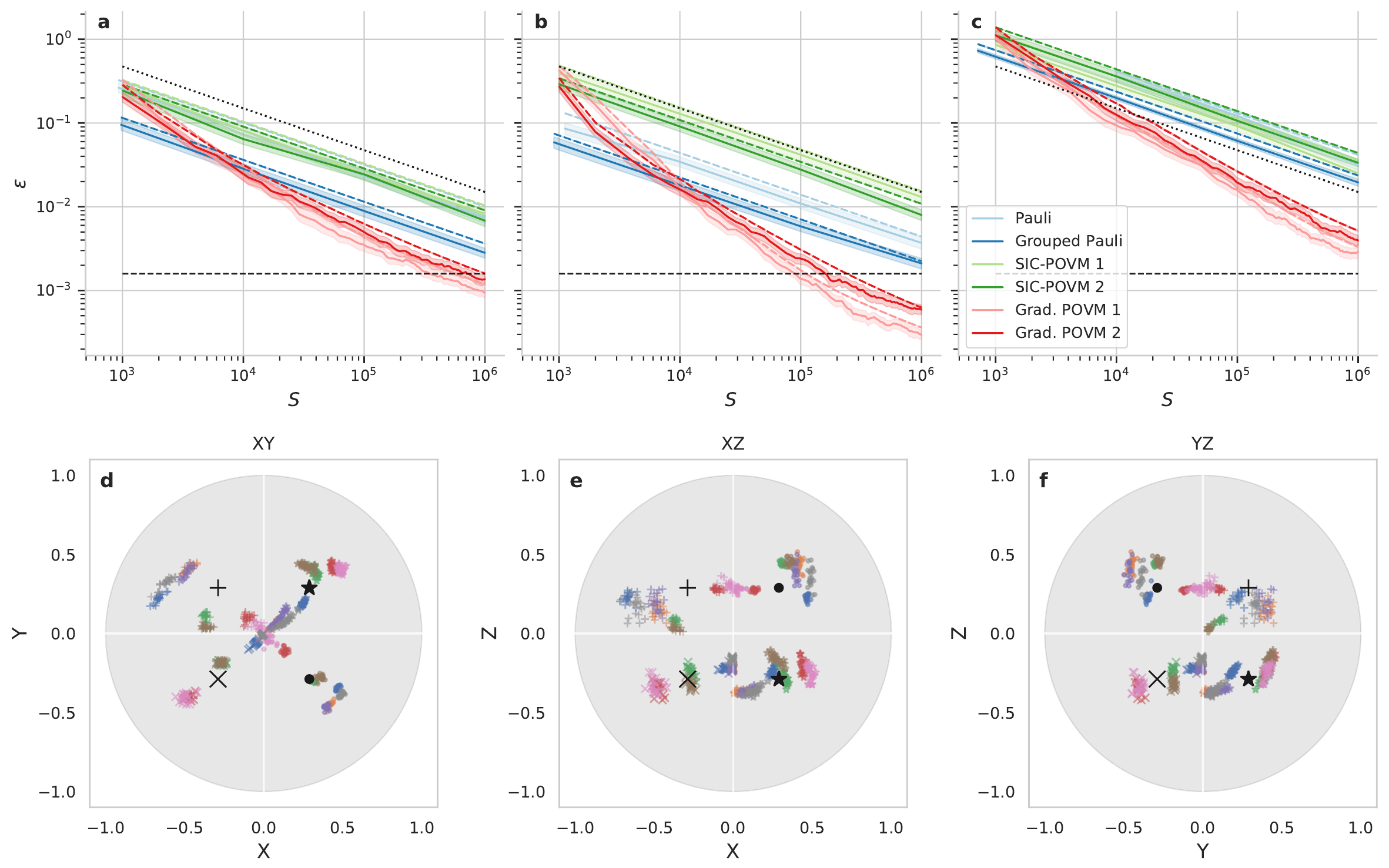}
    \caption{\textbf{Adaptive measurement for energy estimation.} (a-c) solid lines indicate the error in the estimation of the ground energy for selected Hamiltonians as a function of the total number of shots, with a VQE state, for various methods: Pauli, grouped Pauli, as well as the hybrid Monte Carlo without optimization (SIC POVM 1 and 2) and the gradient-optimized method (Grad.~POVM 1 and 2) with two different initial conditions (see App.~\ref{app:sic-povms}). The dashed lines correspond to the estimated statistical error ($\bar{\bar{V}}^{1/2}$ in the case of the POVM methods). The curves show the average error over 100 realisations of the numerical experiments and the shaded areas the $95\%$ confidence interval obtained with bootstrapping. The horizontal dashed line indicates chemical accuracy, while the tilted one illustrates the $S^{-1/2}$ scaling. The three examples are (a) 8-qubit $\rm{H_2}$ with JKMN mapping, (b) 8-qubit $\rm{LiH}$ with BK, and (c) 14-qubit $\rm{H_2O}$ with JKMN.
    (d-f) final POVM effects in the gradient optimization process, when starting from the SIC POVM 2, for a sample of 20 realisations from the data set of (a). Every POVM effect is mapped onto the three-dimensional unit-radius ball in a similar way as how single-qubit states are mapped onto the Bloch ball. In particular, the point $\vec{r} = (r_x, r_y, r_z), \, \vert \vec{r} \vert \leq 1$, is associated with the effect $\Pi(\vec{r}) = (\vert \vec{r} \vert \mathbb{I} + \vec{r} \cdot \vec{\sigma}) / 2$ (note the difference with the Bloch ball representation of quantum states; see App.~\ref{app:povm_implementation}). In the figure, the color indicates the qubit to which an effect corresponds, while the symbol identifies the effect itself among the possible four. The black symbols locate the initial effects, common to all realisations and qubits. Each panel presents the projection of the ball onto a different plane. The clustering of the points with equal color and symbol reveal that all realisations reach approximately the same optimal measurement. However, the result of the optimization is different for every qubit. Moreover, starting with SIC POVM 1 instead leads to a very different measurement (see \cite{videos}).
    }
    \label{fig:energy_estimation}
\end{figure*}

\subsection{Classical gradient estimation for POVM optimization}\label{sec:gradient_estimation}

Modification of the POVM results in a different probability distribution $\lbrace p_\mathbf{m} \rbrace$, as well as different weights $\omega_\mathbf{m}$, and hence potentially different $\mathrm{Var} (\omega_\mathbf{m})$. This can be exploited to devise an adaptive algorithm in which the measurement of $\langle \mathcal{O} \rangle$ is optimized over the space of POVMs, that is, by finding one that minimizes the variance $\mathrm{Var} (\omega_\mathbf{m})$. We now propose a classical post-processing routine to navigate the space of POVMs toward low-variance ones. Essentially, besides using the outcomes obtained with the current POVM to construct an estimation of the target observable, the same set of data is also employed in a classical routine to assess the variance of \textit{other} POVMs that have not previously been implemented on the quantum processor.
Such procedure is explained in detail in the following.

Suppose that we want to evaluate the Monte Carlo variance $\mathrm{Var} (\omega'_\mathbf{r})$ for a new POVM defined in terms of local POVMs with effects $\lbrace \Gamma_r^{(i)}\rbrace$, that is,
\begin{equation}\label{eq:variance_gamma}
\mathrm{Var} (\omega'_\mathbf{r}) = \sum\limits_{\mathbf{r}} {\omega'_{\mathbf{r}}}^2 \mathrm{Tr} [ \rho \Gamma_\mathbf{r} ] - \left( \sum\limits_{\mathbf{r}} {\omega'_{\mathbf{r}}} \mathrm{Tr} [ \rho \Gamma_\mathbf{r} ] \right)^2,
\end{equation}
where the $\omega'_\mathbf{r}$ are given by the $b_{kr}^{(i)}$ matrices corresponding to these local POVMs, and $\Gamma_\mathbf{r} = \bigotimes_{i=1}^{N} \Gamma^{(i)}_{r_i}$. The second term in Eq.~\eqref{eq:variance_gamma} is the squared mean $\langle \mathcal{O} \rangle^2$, which does not depend on the POVM. The first term, i.e.~the second moment of $\omega_\mathbf{r}$ over the probability distribution $\left\lbrace q_\mathbf{r} \equiv \mathrm{Tr} [ \rho \Gamma_\mathbf{r} ] \right\rbrace$, $\left\langle {\omega'_{\mathbf{r}}}^2 \right\rangle_{\lbrace q_\mathbf{r} \rbrace}$, is the one that we must minimize.

Suppose further that we have already run some experiments on the quantum computer with another IC POVM given by the effects $\lbrace \Pi_m^{(i)} \rbrace$. Since this POVM is IC, we can write $\Gamma_{r}^{(i)} = \sum_m d_{rm}^{(i)} \Pi_{m}^{(i)}$, where the $d_{rm}^{(i)}$ are real numbers. Inserting these decompositions into the expression for the second moment, we obtain
\begin{equation}\label{eq:classical_mc2}
\begin{aligned}
\left\langle {\omega'_{\mathbf{r}}}^2 \right\rangle_{\lbrace q_\mathbf{r} \rbrace} &= \sum\limits_{\mathbf{r}} {\omega'_{\mathbf{r}}}^2 \mathrm{Tr} \left[\rho \bigotimes\limits_{i=1}^{N} \left( \sum_{m_i=0}^3 d_{r_{i}m_{i}}^{(i)} \Pi_{m_{i}}^{(i)} \right)\right] \\
&= \sum\limits_{\mathbf{m}} p_{\mathbf{m}} \sum\limits_{\mathbf{r}} \prod\limits_{i=1}^{N} d_{r_{i}m_{i}}^{(i)} {\omega'_{\mathbf{r}}}^2.
\end{aligned}
\end{equation}
This last expression is also calculated in a hybrid Monte Carlo manner. More precisely, we can reuse the strings $\mathbf{m}_1, \ldots, \mathbf{m}_S$ obtained from the measurements on the quantum computer (sampled from the probability distribution $\lbrace p_\mathbf{m} \rbrace$) to estimate the variances of other POVMs by calculating, for each $\mathbf{m}_s$, the corresponding $\sum_{\mathbf{r}} \prod_{i=1}^{N} d_{r_{i}m_{i}}^{(i)} {\omega'_{\mathbf{r}}}^2$ classically. Note, however, that this last sum cannot always be computed efficiently, since it generally contains $4^N$ terms (both positive and negative), and involves products $\prod_{i=1}^{N} d_{r_{i}m_{i}}^{(i)}$ that can scale exponentially in $N$. To ensure the feasibility of the procedure, we use a gradient descent approach for the optimization of the POVMs; in such case, only one of the terms in the product is different from one.

For concreteness, suppose that we use the effects $\lbrace \Pi_m^{(i)} \rbrace$ corresponding to the point $\vec{x}$ in the POVM parameter space (see App.~\ref{app:povm_implementation}) on the quantum computer and obtain $S$ samples with which we can estimate the second moment $\langle \omega_\mathbf{m}^2 \rangle$. We can approximate the partial derivative of the second moment with respect to one of the parameters (for instance, the $k$-th), as $\partial_{x_k} \langle \omega_\mathbf{m}^2 \rangle \approx (\langle {\omega'_\mathbf{r}}^2 \rangle - \langle \omega_\mathbf{m}^2 \rangle) / h$, where $\langle {\omega'_\mathbf{r}}^2 \rangle$ is the estimated second moment corresponding to the POVM the coordinates of which in parameter space $\vec{x}'$ fulfill $x'_k = x_k + h$ and $x'_j = x_j$ for $j \neq k$ (let us denote the corresponding effects by $\lbrace \Gamma_r^{(i)} \rbrace$), and $h \ll 1$.

Since all single-qubit POVM are parametrized independently, $N-1$ of them are identical to the ones already used on the device, that is, $d_{rm}^{(i)} = \delta_{rm}, \, \forall i \neq l$, where $l$ is the qubit the POVM of which depends on the $k$-th parameter. Introducing this expression into Eq.~\eqref{eq:classical_mc2}, we obtain
\begin{equation}\label{eq:classical_mc_grad}
\left\langle {\omega'_{\mathbf{r}}}^2 \right\rangle = \sum\limits_{\mathbf{m}} p_{\mathbf{m}} \sum\limits_{r_l = 0}^{3} d_{r_{l}m_{l}}^{(l)} {\omega'}^2_{(m_1,\ldots, m_{l-1}, r_{l}, m_{l+1}, \ldots, m_N)}.
\end{equation}
Using this method, all the partial derivatives can be calculated using classical post-processing, in polynomial time, of the same samples $\mathbf{m}_1, \ldots, \mathbf{m}_S$ obtained from the quantum computer. Once the gradient has been estimated, we can identify a new POVM with smaller expected variance than the previous one. We detail the gradient-based optimization used in this work in App.~\ref{app:gradient_descent}.

\begin{figure*}[t!]
    \centering
    \includegraphics[width=\textwidth]{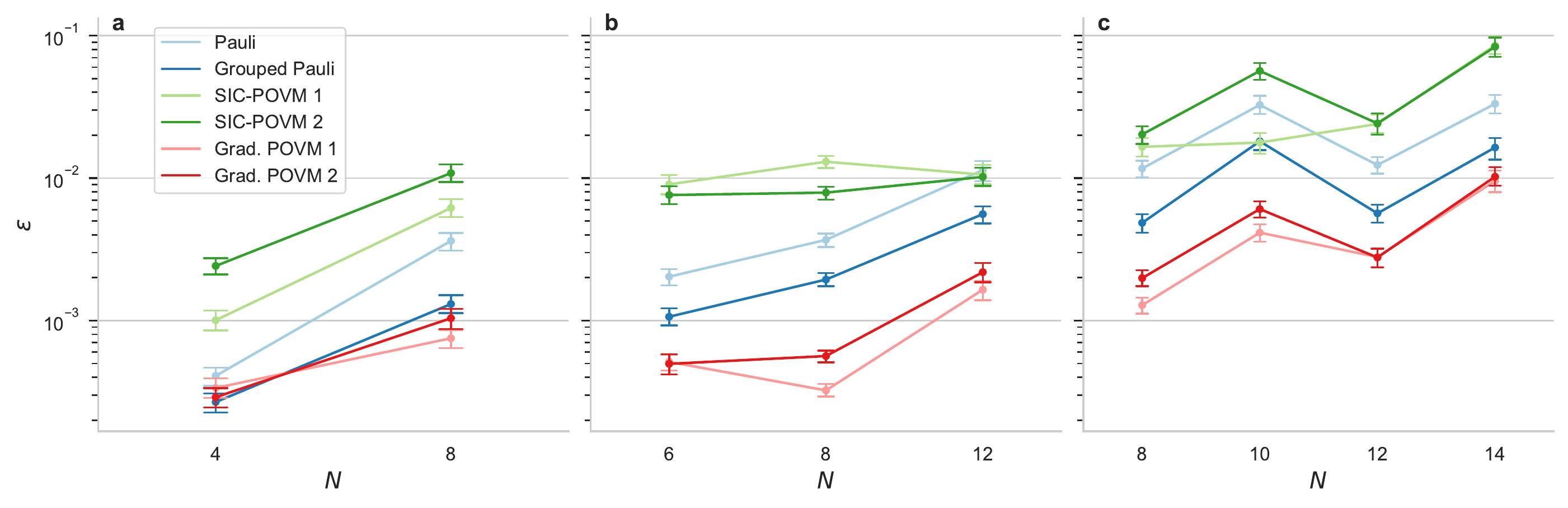}
    \caption{\textbf{Performance of the adaptive measurement scheme.} Error in the estimation of the ground energy of the three molecules with the BK fermion-to-qubit mapping. Every plot compares the results for one molecule ($\rm{H_2}$ (a), $\rm{LiH}$ (b), and $\rm{H_2O}$ (c)) with different measurement methods, with a total of $S = 10^6$ shots (for the Pauli and grouped Pauli methods, we use the same number of shots $\lfloor 10^6 / K \rfloor$ on every Pauli string, so the total number of shots is in fact $S = K \lfloor 10^6 / K \rfloor$; this represents a deficit of at most $0.1\%$ in the total number of shots in the examples considered). The ground state is approximated by optimising a VQE ansatz. The estimation error is the absolute difference between the simulation results and the exact value for the optimized ansatz. The points represent the average error over 100 realisations and the error bars show a $95\%$ confidence interval obtained using bootstrapping. For $\rm{H_2}$, our algorithm offers little improvement, but the difference in performance becomes clearer with the other two molecules. Note that the two initial POVMs yield slightly different results, with SIC POVM 1 generally outperforming SIC POVM 2. We also note that, in the cases involving more qubits, such as the 14-qubit $\rm{H_2O}$ molecule with the BK mapping, the measurement optimization has not fully converged for $S = 10^{6}$ shots, so the difference with respect to the other methods is expected to increase for larger $S$, potentially reaching chemical accuracy earlier. The results with other fermion-to-qubit mappings are available in App.~\ref{app:other_experiments}.}
    \label{fig:energy_errors}
\end{figure*}

\subsection{On-the-fly optimization}\label{sec:on_the_fly}

An important aspect of the algorithm is that we do not need to first optimize the POVM (until it reaches a small-enough variance) before starting to estimate the expected value of the observable. The intermediate POVMs used in the process are also IC, so they can be used for the estimation of $\langle \mathcal{O} \rangle$ as well. The strategy is to use the intermediate mean values obtained with every fixed choice of the POVM to calculate a weighted average. As we will show below, the latter is designed in a way that minimizes the resulting variance in the overall estimation. The whole procedure can be carried out iteratively as the algorithm progresses, thus effectively making use of all measurement results obtained during the intermediate POVM optimization steps for the reconstruction of~$\langle \mathcal{O} \rangle$.

For the sake of clarity, let us first consider the situation in which we use $T$ different POVMs, each with $S_t, t \in [1, T]$ shots (i.e., statistical samples or repetitions of the measurement protocol), and we produce $T$ different estimations $\bar{\mathcal{O}}_t$ using Eq.~\eqref{eq:estimated_mean} for each of the individual POVM choices. These estimations are, in fact, random variables with the same mean $\langle \mathcal{O} \rangle$ but different variance, $\textrm{Var} ( \bar{\mathcal{O}}_t ) = \textrm{Var} (\omega_\mathbf{m}^{(t)} ) / S_t$.

Let us now define a new random variable, $\bar{\bar{\mathcal{O}}}_{2} (\alpha) = \alpha \bar{\mathcal{O}}_2 + (1 - \alpha) \bar{\mathcal{O}}_1$, where $\alpha \in (0, 1)$ is a parameter of our choice, the mean of which is $\langle \bar{\bar{\mathcal{O}}}_{2} (\alpha) \rangle = \alpha \langle \bar{\mathcal{O}}_2 \rangle + (1 - \alpha) \langle \bar{\mathcal{O}}_1 \rangle = \langle \mathcal{O} \rangle$ for any value of $\alpha$. Its variance is, however, a function of $\alpha$. Since the $\bar{\mathcal{O}}_t$ are all independent random variables, we have $\textrm{Var} (\bar{\bar{\mathcal{O}}}_{2} (\alpha)) = \alpha^2 \textrm{Var} (\bar{\mathcal{O}}_2) + (1 - \alpha)^2 \textrm{Var} (\bar{\mathcal{O}}_1)$. While we do not know the actual values of the variances $\lbrace \textrm{Var} (\bar{\mathcal{O}}_t) \rbrace$, we can estimate them from the data; let us refer to the corresponding estimation of $\textrm{Var} (\bar{\mathcal{O}}_t)$ with the symbol $\bar{V}_t$. This allows us to estimate the variance of $\bar{\bar{\mathcal{O}}}_{2} (\alpha)$ as $\textrm{Var} (\bar{\bar{\mathcal{O}}}_{2} (\alpha)) \approx \alpha^2 \bar{V}_2 + (1 - \alpha)^2 \bar{V}_1$. This quantity is minimized for $\alpha_{\mathrm{opt}} = \bar{V}_1 / (\bar{V}_1 + \bar{V}_2)$, yielding an estimated variance $\textrm{Var} (\bar{\bar{\mathcal{O}}}_{2} (\alpha_{\mathrm{opt}})) \approx \bar{\bar{V}}_2 \equiv \bar{V}_1 \bar{V}_2 / (\bar{V}_1 + \bar{V}_2)$, which is smaller than $\bar{V}_1$ and $\bar{V}_2$. Thus, we have combined two different estimators of $\langle \mathcal{O} \rangle$ to produce a new one with smaller statistical error. Next, define the random variable $\bar{\bar{\mathcal{O}}}_{3} (\alpha') = \alpha' \bar{\mathcal{O}}_3 + (1 - \alpha') \bar{\bar{\mathcal{O}}}_{2} (\alpha_{\mathrm{opt}})$. Using the same arguments as before, the value of $\alpha'$ minimising the variance of $\bar{\bar{\mathcal{O}}}_{3}$ is $\alpha'_{\mathrm{opt}} = \bar{\bar{V}}_2 / (\bar{\bar{V}}_2 + \bar{V}_3)$, giving $\textrm{Var} (\bar{\bar{\mathcal{O}}}_{3} (\alpha_{\mathrm{opt}})) \approx \bar{\bar{V}}_3 \equiv \bar{\bar{V}}_2 \bar{V}_3 / (\bar{\bar{V}}_2 + \bar{V}_3)$. The process can be iterated for the $T$ estimators.

The above procedure can be recast in terms of an iterative algorithm as follows:
\begin{itemize}
\item[1.] Initialize two variables, $\bar{\bar{\mathcal{O}}}$ and $\bar{\bar{V}}$, such that $\bar{\mathcal{O}}_{1} \rightarrow \bar{\bar{\mathcal{O}}}$ and $\bar{V}_1 \rightarrow \bar{\bar{V}}$.
\item[2.] At the end of each iteration $t \in (2, \ldots, T)$ of the POVM optimization, update them as $(\bar{\mathcal{O}}_{t} \bar{\bar{V}} + \bar{\bar{\mathcal{O}}} \bar{V}_{t}) / (\bar{\bar{V}} + \bar{V}_{t}) \rightarrow \bar{\bar{\mathcal{O}}}$ and $\bar{\bar{V}} \bar{V}_{t} / (\bar{\bar{V}} + \bar{V}_{t}) \rightarrow \bar{\bar{V}}$.
\end{itemize}
At any point along the process, we have an estimated mean $\bar{\bar{\mathcal{O}}}$ with estimated standard error $\bar{\bar{V}}^{1/2}$ that minimizes the overall error of the input data and can be easily updated with new ones. It is important to stress that this iterative mixing of the outcomes is unbiased, as we prove in App.~\ref{app:sequential_vs_one-step}.

\section{Numerical simulations}\label{sec:results}

\begin{figure}[t]
    \centering
    \includegraphics[width=\columnwidth]{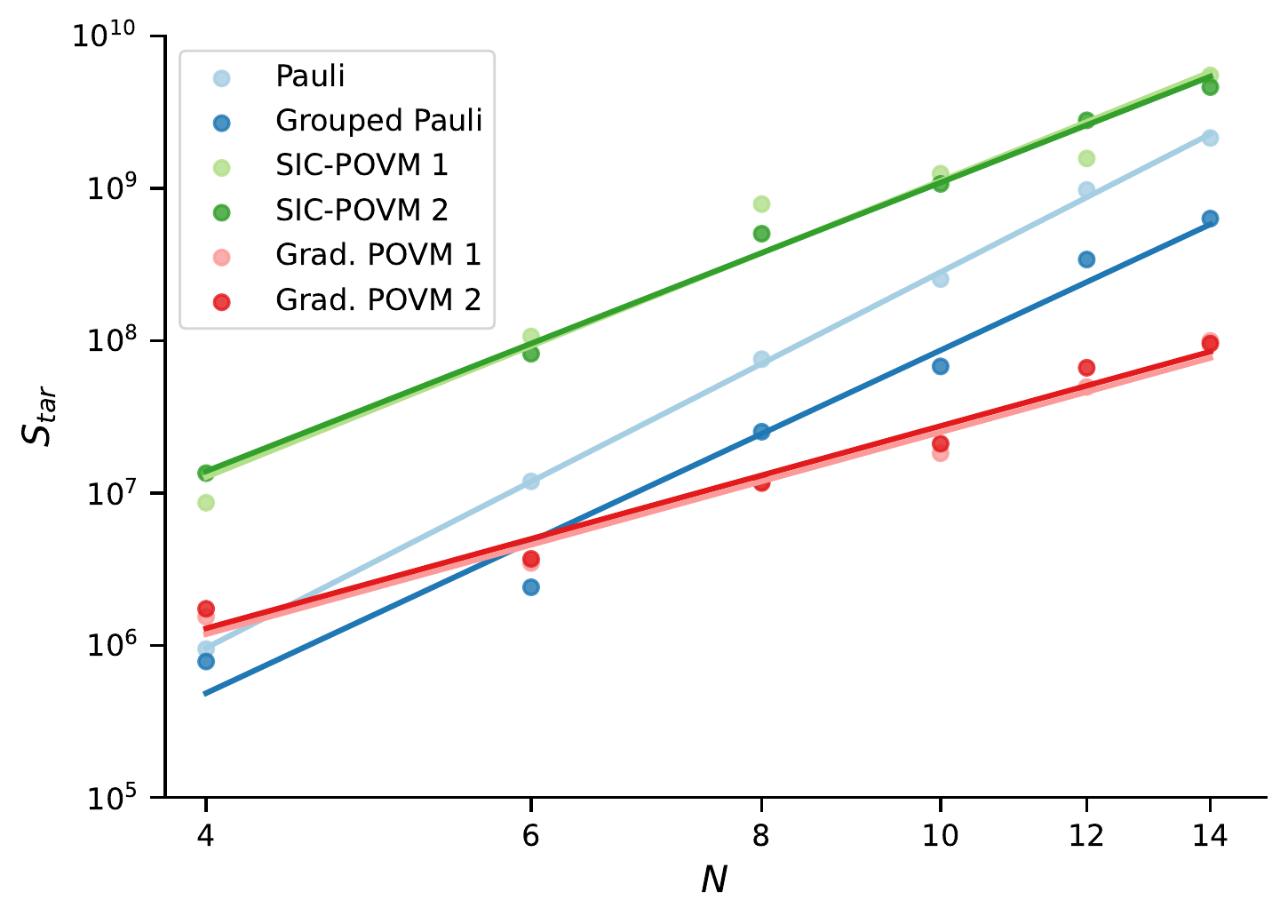}
    \caption{\textbf{Scaling of the adaptive measurement scheme.}
    Number of shots $S_\mathrm{tar}$ required to achieve a target estimated error of $\epsilon_\mathrm{tar} = 0.5$ mHa for $\rm{H}$ chains as a function of the number of qubits $N$. The qubit Hamiltonian is obtained using the JKMN mapping. For each method and molecule, we use up to $S_\mathrm{lim} \approx 10^6$ runs, as in Fig.~\ref{fig:energy_errors}. If the average estimated error with $S_\mathrm{lim}$ shots, $\epsilon_\mathrm{lim} = \langle \bar{\bar{V}}^{1 / 2} \rangle$, where $\langle \cdot \rangle$ represents the average over realisations, is still larger than $\epsilon_\mathrm{tar}$, we estimate the required number of shots needed to reach it by assuming a scaling $\epsilon \sim S^{-1/2}$, that is, we use $S_\mathrm{tar} = S_\mathrm{lim} \epsilon^2_\mathrm{lim} / \epsilon^2_\mathrm{tar}$. While this procedure saves us considerable computing time, it also overestimates the number of measurements needed by our algorithm: indeed, the convergence of our method to $\epsilon_\mathrm{tar}$ is faster than $\epsilon \sim S^{-1/2}$ unless it has already converged to the optimal POVM (see Fig.~\ref{fig:energy_estimation}). Thus, these results are to be regarded as an upper bound to the total measurement cost of the learning POVM method. The curves depict least squares fits to the data with functions of the form  $S_\mathrm{tar} = a N^b$. The corresponding values of the exponent $b$ for each method are reported in Table \ref{tab:scaling_exponents}. Note that the values found for the Pauli and grouped Pauli methods are consistent with the ones reported in Ref.~\cite{Huggins2019}. Moreover, the performance of our algorithm is similar to that of the state-of-the-art method proposed in Ref.~\cite{Huggins2019}, especially for the lower values of $N$, for which the overestimation of $S_\mathrm{tar}$ is less significant.}
    \label{fig:scaling_plot}
\end{figure}

In this section, we present the results of the numerical experiments that are run to test the feasibility and performance of our algorithm. Section \ref{sec:energy_measurement_learning} is aimed at illustrating the effect of the adaptive measurement. Section \ref{sec:performance_and_scaling} presents a more in-depth analysis of the performance. Finally, in Sec.~\ref{sec:ic_data} we demonstrate an important feature of our approach: the IC data used for the estimation of the energy can be reused for other purposes. All the data used in this manuscript are available on Zenodo \cite{dataset}. The source code used to generate the results is available online \cite{sourcecode}.

\subsection{Energy measurement learning}
\label{sec:energy_measurement_learning}
We start by measuring the ground-state energy of the
$\rm{H_2}$, $\rm{LiH}$ and $\rm{H_2O}$ molecules. For the characterization of each system, we use different numbers of molecular orbitals. The basis set used for $\rm{H_2}$ is 6-31G~\cite{dithcfield1971self,hehre1972self,hariharan1973influence,hariharan1974accuracy} leading to 8 spin orbitals, while for the case of $\rm{LiH}$ and $\rm{H_2O}$ we used STO3G~\cite{Hehre1969self,Collins1976self} basis set leading to 12 and 14 spin orbitals, respectively. We use the Bravyi-Kitaev (BK)~\cite{bk2002}, the Jiang et al. (JKMN)~\cite{Jiang2020optimalfermionto}, the Jordan-Wigner (JW), and the parity~\cite{Bravyi17} mapping transformations. The latter has an intrinsic property, deriving from spin up and spin down electron conservation, that reduces the number of qubits required by two~\cite{Bravyi17}. We also leverage different symmetries present in each system to reduce further the qubit count~\cite{Bravyi17}. For the case of $\rm{LiH}$ and $\rm{H_2O}$ we also freeze the core orbitals allowing us to exclude another two spin orbitals from our calculation (refer to the table in Fig.~\ref{fig:all_results} for more details on the Hamiltonians and qubit reductions considered). Each of these molecular Hamiltonians is mapped into qubits using one or more of the aforementioned techniques, hence producing several qubit Hamiltonians with varied number of qubits, which are then used to simulate the energy measurement process in a VQE experiment near convergence.

We proceed as follows.
First, for each qubit Hamiltonian $H$, we numerically approximate the ground state with a hardware-efficient ansatz $\vert \psi(\vec{\theta}) \rangle$ introduced in Ref.~\cite{Kandala17}. This generates a trial wave function by combining repetitive layers of single qubit $R_y$ gates and entangling blocks composed of two-qubit operations [controlled-NOT (CNOT) gates]. The single qubit rotations are parametrized with a set of angles (also known as variational parameters) that are iteratively updated, with the help of a classical optimization routine, in order to minimize the energy expectation value. Once we have the optimal parameters for which the variational form $\vert \psi(\vec{\theta}_{\mathrm{opt}}) \rangle$ approximates the ground-state wave function, we calculate the corresponding exact expected energy $\langle E \rangle = \langle \psi(\vec{\theta}_{\mathrm{opt}}) \vert H \vert \psi(\vec{\theta}_{\mathrm{opt}}) \rangle$.

We then simulate different energy evaluation methods as a function of the number of state preparations (shots) $\bar{E}(S)$, and compute the corresponding errors $\vert \bar{E}(S) - \langle E \rangle \vert$.
We also calculate the estimated statistical error for each approach, that is, the estimated error when the exact value $\langle E \rangle$ is not available (for the gradient-descent algorithm, this error is given by $\bar{\bar{V}}^{1/2}$ as defined in Sec.~\ref{sec:on_the_fly}). These quantities are depicted in Fig.~\ref{fig:energy_estimation} (a-c) for three selected examples.

The effect of the measurement learning results in the error decreasing faster than $S^{-1/2}$, especially for small $S$. This is a consequence of the fact that, after each batch of runs, the next POVM used in the sequence is in principle more efficient (i.e.\ leads to a smaller variance) than the previous one. Importantly, even if the starting efficiency is lower than that of other methods, our algorithm eventually takes over and reaches better accuracy at lower costs. Moreover, as we discuss in detail in the next subsection, even the use of Eq.~\eqref{eq:estimated_mean} with the initial POVM without optimization tends to give better performance than with the Pauli and the grouped Pauli methods, as the size of the problem increases. The results also reveal that $\bar{\bar{V}}^{1/2}$, as introduced in Sec.~\ref{sec:on_the_fly}, gives the correct estimation of the statistical error in the evaluation of the energy~\footnote{Note that the small difference between the measured and the estimated errors is entirely due to the definition of error used, i.e.~$\langle \vert \bar{E}(S) - \langle E \rangle \vert \rangle$. Indeed, using $\sqrt{\langle ( \bar{E}(S) - \langle E \rangle )^2 \rangle}$ instead leads to a perfect overlap between both (not shown).}.

The learning process is also illustrated in Fig.~\ref{fig:energy_estimation} (d-f), where we depict graphically the result of the optimization in terms of a geometric representation of the effects akin to the Bloch sphere for single-qubit states (see App.~\ref{app:povm_implementation} for details). We only include the results for one example in the paper, but the results for all the Hamiltonians analysed in this work, as well as their animated version, are available online \cite{videos}. Interestingly, while the optimization eventually converges and different realisations with the same initial condition lead to the same minimum (modulo small fluctuations), the two initial conditions considered here (see App.~\ref{app:sic-povms}) result in different optimal POVMs with slightly different performance. This suggests the potential existence of better initial conditions than those explored here. This subject will be considered in future work.

\begin{table}[t]
    \centering
    \begin{ruledtabular}
\begin{tabular}{l|llll}
{} & \multicolumn{4}{c}{Mapping} \\
Method          &   JW            & Parity           & BK            & JKMN  \\
\hline
Pauli         &  $6.4 \pm 0.4$ &  $6.3 \pm 0.3$ &  $6.8 \pm 0.5$ &  $6.2 \pm 0.1$ \\
grouped Pauli &  $5.1 \pm 0.5$ &  $5.6 \pm 0.4$ &  $6.3 \pm 0.5$ &  $5.7 \pm 0.5$ \\
SIC POVM 1    &  $5.1 \pm 0.6$ &  $5.8 \pm 0.7$ &  $5.9 \pm 0.6$ &  $4.9 \pm 0.5$ \\
SIC POVM 2    &  $5.8 \pm 0.7$ &  $5.3 \pm 0.6$ &  $4.4 \pm 0.5$ &  $4.8 \pm 0.2$ \\
Grad. POVM 1  &  $3.0 \pm 0.3$ &  $4.5 \pm 0.6$ &  $4.8 \pm 0.6$ &  $3.3 \pm 0.3$ \\
Grad. POVM 2  &  $3.2 \pm 0.4$ &  $4.4 \pm 0.6$ &  $4.7 \pm 0.5$ &  $3.3 \pm 0.3$ \\
\end{tabular}
\end{ruledtabular}

    \caption{\textbf{Scaling exponents of the adaptive measurement scheme.}
    The points in Fig.~\ref{fig:scaling_plot}, as well as their counterparts using other fermion-to-qubit mappings, are fitted to a function of the form $S_\mathrm{tar} = a N^b$. The table contains the corresponding exponents. The exponent $b \approx 6$ of the Pauli method, as well as the mild reduction $b \approx 5.6$ offered by grouped Pauli, are consistent with the values reported in Ref.~\cite{gonthier2020identifying} for other molecules. The POVM-based method without optimization already outperforms these results, with $b \approx 4.8$ using the JKMN mapping~\cite{Jiang2020optimalfermionto}. The adaptive strategy results in a considerably smaller exponent $b \approx 3.3$. Interestingly, a similar scaling is achieved also for the JW mapping; while this mapping leads to Pauli strings with weight of $\mathcal{O}(N)$, the adaptive strategy is able to find POVMs for which the measurement process is efficient.}
    \label{tab:scaling_exponents}
\end{table}

\subsection{Performance and scaling}
\label{sec:performance_and_scaling}
While the previous results illustrate the working principles of the algorithm with three molecular Hamiltonians, we now turn our attention toward the analysis of its performance. In Fig.~\ref{fig:energy_errors} and in App.~\ref{app:other_experiments}, we collect the errors of similar estimations for several other Hamiltonians corresponding to the same molecules (under different qubit-reduction schemes) for a total number of measurements $S \approx 10^{6}$, from which it can be seen that our algorithm is advantageous in almost all cases, and particularly for $\rm{LiH}$ and $\rm{H_2O}$. Note that, since our algorithm is adaptive and the error decreases faster than $S^{-1/2}$, in contrast to the other methods, the advantage of the former would potentially increase for larger $S$. This is especially the case for the larger problems, for which the POVM-learning algorithm is further from convergence ---and the error in the energy from chemical accuracy--- at $S = 10^6$ shots.

In order to study the performance of the algorithm for larger Hamiltonians, we analyse the number of measurements required to reach an accuracy of 0.5 mHa as a function of the number of qubits for hydrogen chains with increasing number of atoms; arguably, this figure of merit is more informative of the usefulness of the approach in real applications, in which one is interested in determining the ground-state energy within some fixed accuracy, rather than obtaining the best performance for a fixed number of shots.

Due to limitations in computational power, we run our simulations for a limited number of measurements and extrapolate the total number required for such precision (see Fig.~\ref{fig:scaling_plot} for results using the JKMN mapping and caption for a detailed explanation). Even though this method overestimates the actual number of shots needed by our algorithm, we see a considerable improvement with respect to the Pauli and grouped Pauli methods. Interestingly, the bare hybrid quantum-classical Monte Carlo method without optimization, despite yielding higher errors for the small sizes considered here, also shows a more favorable scaling than the former methods.

To provide a more quantitative evaluation, we further fit each set of results into a function of the form $a N^b$. We report the corresponding values of the exponent $b$, also including those for other fermion-to-qubit mappings, in Table \ref{tab:scaling_exponents}. Note that, while other mappings are added for completeness, the optimal performance of our algorithm is expected with the mapping from Ref.~\cite{Jiang2020optimalfermionto} (see Sec.~\ref{sec:monte_carlo}), as confirmed by the results. Importantly, we can see that our method thus benefits from two improvements: the Monte Carlo approach results in a considerable reduction in the exponent, followed by a second scaling improvement stemming from the learning strategy. The result is an overall efficiency comparable to state-of-the-art methods~\cite{Huggins2019, Jiang2020optimalfermionto}.

\begin{figure*}[t]
    \centering
    \includegraphics[width=.85\textwidth]{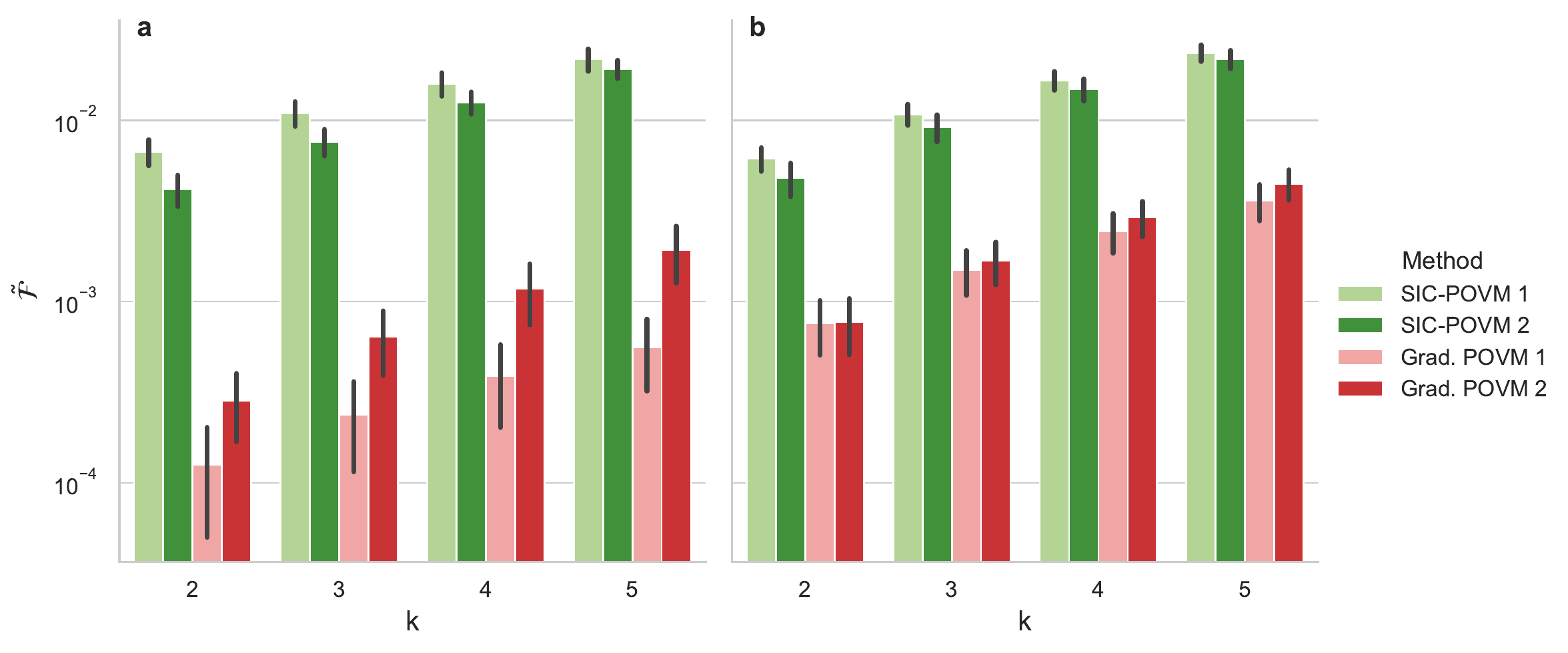}
    \caption{\textbf{Reduced tomography from energy estimation data.} Average $k$-wise state infidelity $\tilde{\mathcal{F}}$, as a function of $k$, of the tomographic reconstructions obtained from the same data used for the estimation of the ground energy of (a) 6-qubit $\rm{LiH}$ with BK mapping and (b) 8-qubit $\rm{H_2}$ with JKMN mapping, with four different methods. The green bars represent the results for the SIC POVM 1 (lighter) and 2 (darker)  (see App.~\ref{app:sic-povms}) without optimization, while the red bars show the results for the gradient descent-optimized POVM 1 (lighter) and 2 (darker) data starting with the same initial condition. Every realisation involves a total of $S = 10^5$ measurements. The bars indicate the average over 20 realisations, while the error bars the standard deviations.}
    \label{fig:tomograhpy}
\end{figure*}

\subsection{Exploiting informationally complete data}
\label{sec:ic_data}
Further numerical experiments demonstrate that the IC data collected for the estimation of the energy can indeed be reused for other purposes. As explained in Sec.~\ref{sec:monte_carlo}, the same IC outputs can be post-processed to calculate any expectation value of our choice, the only limitation being that, as it is reasonable to expect, the optimization procedure targeting a particular observable may worsen the estimation of other specific quantities. In what follows, rather than focusing on particular additional observables, we consider an arguably more costly task: state tomography. More precisely, we address the reconstruction of all the $k$-qubit density operators in the system for all $k \leq K$. Reduced tomography has recently attracted some interest in the quantum information literature for diverse purposes~\cite{Jena2019,Yen2019b,Huggins2019,Gokhale2019,Crawford2019,zhao2019measurement,paini2019approximate,BonetPhysRevX2020,CotelrPRL2020,GGP_pairwise_2020,garcia2021experimentally}.

We thus proceed in a similar manner as in the previous subsections. We approximate the ground states by training VQE ans\"atze and then estimate the energy using the adaptive algorithm. The resulting data is then used to reconstruct all the $k$-qubit reduced density matrices using likelihood maximisation. In particular, for every subset of $k$ qubits in the system, we marginalise the outcomes over the subset and then use the algorithm introduced in Ref.~\cite{vrehavcek2007diluted} to reconstruct the density operator.
Since we must integrate IC data from $T$ different POVMs in the likelihood maximisation procedure, we define a collective POVM with $T \times 4^N$ effects $\lbrace \Xi_{(t, \mathbf{m})} = \Pi_\mathbf{m}^{(t)} S_t / S, t \in [1, T] \rbrace$, where the index $t$ indicates the POVM optimization step, $S_t$ represents the number of measurements carried out in iteration $t$, and $S = \sum_t S_t$~\footnote{Note that this definition is consistent with the number of shots $S_t$ imposed at each iteration $t$: for any state $\rho$, the expected number of outcomes corresponding to effects $\lbrace \Xi_{(t, \mathbf{m})} \rbrace$ with a particular $t$, given a total number of shots $S = \sum_t S_t$, is $S \sum_\mathbf{m} \mathrm{Tr} [\rho \Xi_{(t, \mathbf{m})} ] = S \mathrm{Tr} [\rho \mathbb{I} S_t / S ] = S_t$.}. Once a $k$-qubit density matrix $\rho_\mathrm{tomo}$ is reconstructed, we compute its infidelity $\tilde{\mathcal{F}}(\rho_{\text{\tiny{tomo}}}, \rho_{\text{\tiny{exact}}}) = 1- \mathcal{F}(\rho_{\text{\tiny{tomo}}}, \rho_{\text{\tiny{exact}}})$,
(where
$ \mathcal{F}(\rho_{\text{\tiny{tomo}}}, \rho_{\text{\tiny{exact}}}) = \mathrm{Tr} [\sqrt{\sqrt{\rho_{\text{\tiny{tomo}}}}\rho_{\text{\tiny{exact}}}\sqrt{\rho_{\text{\tiny{tomo}}}}}]^2$ is the quantum fidelity) with respect to the exact one $\rho_{\text{\tiny{exact}}}$ (obtained by tracing out all other qubits in the trained VQE ansatz).

In Fig.~\ref{fig:tomograhpy}, we show the resulting average $k$-wise infidelity for the ground states of two molecules, $\rm{H_2}$ and $\rm{LiH}$, as a function of $k$, with and without gradient-based POVM optimization. We note that the density matrices can be reconstructed with high fidelity from the same data that was used for the estimation of the energy. Moreover, the comparison between these two methods reveals that the optimization of the POVM with respect to the precision in the estimation of energy also improves the fidelity of the reconstructed density matrices by up to an order of magnitude. In all cases, however, the infidelity increases with $k$, as expected.

\section{Discussion and conclusions}
We introduce an algorithm for efficient observable estimation that exploits informationally complete generalized quantum measurements integrating three important components: a hybrid quantum-classical Monte Carlo, an efficient method to navigate POVM space toward low-variance measurements, and a recipe to combine different estimations of the observable of interest. The result is a procedure in which an optimized measurement of an operator average is learnt in an adaptive fashion with no measurement overhead. Consequently, the overall measurement cost is drastically reduced with respect to the initial POVM considered. This is particularly interesting for real applications, considering that the initial SIC POVMs used already offer a significant improvement over other widely used methods, such as grouped Pauli. Importantly, the method does not require any exponentially scaling classical or quantum computations, although it does involve a modest polynomial classical overhead.

We have illustrate the potential of the approach with several proof-of-principle numerical experiments by reconstructing the ground-state energies of several molecular Hamiltonians.
Importantly, our simulations suggest that this adaptive method exhibits scaling performance comparable to those of the most efficient measurement-reduction techniques in the current literature. While confirmation of this calls for a more thorough analysis and simulations, possibly including more general operators than molecular Hamiltonians, it is also important to point out that there is still substantial room for improvement in our algorithm, especially in the parametrisation of the POVMs and in the gradient-descent-based update schedule. It might also be interesting to investigate, in a future work, the potential use of classical machine-learning methods to enhance the measurement-adaptation step~\cite{quek_adaptive_2021}.

Our algorithm also offers some other intrinsic advantages. Being completely agnostic to the nature of the qubit Hamiltonian, and not inspired by quantum chemistry but by quantum information alone, the proposed procedure may find interesting applications beyond VQE calculations. The method is also formally exact, as no approximations are made at any point, except for using the estimated variances as proxies of the actual ones. Moreover, the informationally complete data produced during the measurement process for a particular observable can in principle be reused to calculate many other properties of the underlying quantum state, including its tomographic reconstruction. We provide evidence of the feasibility of this prospect by performing high-fidelity reduced state tomography with no additional measurements.

In this paper, we have only consider the task of estimating a given observable for a fixed quantum state. This typically represents a single step of, e.g., a VQE calculation. In perspective, one could, however, easily integrate our proposed method as a subroutine of the whole ansatz-optimization method. In such case, it might actually be helpful to use the optimal POVM from the previous VQE step, or a slight modification of it, as the starting point of the measurement optimization on the updated ansatz, given that the trial wave function should undergo relatively small changes between consecutive iterations. While this stands as a hypothesis for now, it might reduce the average number of steps required to adjust the POVM settings, hence leading to an even larger reduction in the measurement costs associated with the overall ansatz-optimization process.

Admittedly, our contribution presents a drawback: it requires twice as many qubits. However, it is important to discuss what this entails in practice. The ancillary qubits used for the implementation of the POVM are initialised in the ground state, along with the rest of the qubits in the device, but no operations are applied on them until the measurement stage. Hence, the algorithm introduced here does not require the entanglement of $2N$ qubits throughout the whole computation, which would amplify the detrimental effect of decoherence. Instead, the additional $N$ qubits should be regarded as nothing other than part of the measurement apparatus. In addition, we note that the method offers a significant advantage over a simpler use of the additional qubits, such as, e.g., executing two grouped Pauli iterations in parallel, which, albeit cutting the total run time by up to a factor of 2, would not improve the scaling of such method.

Finally, it is worth discussing some relevant aspects regarding its implementation on real hardware, especially on near-term quantum computers. On the one hand, in devices with limited connectivity, additional SWAP gates may be required in order to enable the interaction between system and ancillary qubits. Importantly, the topology of most currently existing platforms enables the additional SWAP gates, when needed, to be parallelized in such a way that the measurement circuit preserves its size independence. This highlights a favorable aspect of the algorithm: since only a constant-depth measurement circuit is required (namely, the application of a two-qubit gate instead of a single-qubit one, and perhaps some SWAP gates if the connectivity requires so, for every system qubit), the measurement process itself is not expected to introduce significant decoherence effects with respect to applying standard Pauli measurements.

Moreover, the commonly used readout noise-reduction techniques, such as the algorithms integrated in Qiskit or any other error-mitigation strategies that would be used for basic Pauli measurements, can be used here to correct the outcome statistics. While a proper assessment of the performance of the method under real noise conditions, as well as possible specific noise-mitigation strategies, is beyond the scope of this work, these considerations suggest that the ideas introduced in this work can play an important role in enabling the first useful applications of quantum computing for quantum chemistry, so far estimated to require prohibitive computing times.

\section*{Acknowledgements}
G.G.-P., M.A.C.R., B.S, and S.M. acknowledge financial support from the Academy of Finland via the Centre of Excellence program (Project No. 336810 and Project No. 336814). The computer resources of the Finnish IT Center for Science (CSC) and the Finnish Grid and Cloud Infrastructure (FGCI) project (Finland) are acknowledged. S.M. and G.G.-P. acknowledge support from the emmy.network foundation under the aegis of the Fondation de Luxembourg. G.G.-P. acknowledges support from the Academy of Finland via the Postdoctoral Researcher program (Project No. 341985). I.T. acknowledges support from the Swiss National Science Foundation through the National Centers of Competence in Research (NCCR) SPIN. IBM, the IBM logo, and ibm.com are trademarks of International Business Machines Corp., registered in many jurisdictions worldwide. Other product and service names might be trademarks of IBM or other companies. The current list of IBM trademarks is available at \url{https://www.ibm.com/legal/copytrade}.

\appendix

\section{POVM parametrisation and circuit implementation}\label{app:povm_implementation}

As stated in the main text, the algorithm relies on parametrized, informationally complete POVMs implemented through the application of two-qubit unitaries with ancillary qubits, followed by projective measurements on the computational basis. To explain the parametrisation used in this work, it is easier to start by identifying the POVM characterizing one such measurement when applying an arbitrary unitary gate $U$ between some system qubit $q$ in state $\rho$ and an ancilla $a$ in state $\ket{0} \bra{0}$. Since the two qubits are eventually measured projectively in the computational basis, there are four possible outcomes $(b_q, b_a)$ with $b_q \in \lbrace 0, 1 \rbrace$ (and similarly for $b_a$). Each outcome occurs with probability $p_{(b_q, b_a)} = \bra{b_q b_a} U \rho \otimes \ket{0} \bra{0} U^{\dagger} \ket{b_q b_a}$. Writing $U = \sum_{ijkl} u_{kl}^{ij} \ket{ij} \bra{kl}$, this expression becomes $p_{(b_q, b_a)} = \sum_{k k'} u_{k0}^{b_q b_a} ( u_{k'0}^{b_q b_a} )^{*} \bra{k} \rho \ket{k'} = \mathrm{Tr} \left[ \ket{\pi_{(b_q, b_a)}} \bra{\pi_{(b_q, b_a)}} \rho \right]$, where we have defined $\ket{\pi_{(b_q, b_a)}} = \sum_k ( u_{k0}^{b_q b_a} )^{*} \ket{k}$. Hence, the corresponding POVM is given by the set of effects $\lbrace \Pi_i = \ket{\pi_i} \bra{\pi_i}, \, i \in [0, 3] \rbrace$, where we have relabelled the outcomes using $i = 2 b_q + b_a$.

The previous calculation suggests our strategy for the POVM parametrisation: parametrize the unitary $U$, and compute the resulting POVM. The following observations are important. Firstly, not all the components $u_{kl}^{b_q b_a}$ are relevant for the measurement, as the initial state of the ancilla deems those with $l = 1$ irrelevant (provided that $U$ is unitary). Secondly, global phases on $\ket{\pi_i}$ have no effect on the resulting operator $\Pi_i$, so we are free to set $u_{00}^{b_q b_a} \in \mathbb{R}$. Thirdly, $U^{\dagger} U = \mathbb{I}$ implies $\sum_{b_q b_a} ( u_{k0}^{b_q b_a} )^{*} u_{k'0}^{b_q b_a} = \delta_{k k'}$, that is, $u_{00}^{b_q b_a}$ and $u_{10}^{b_q b_a}$ are the components of two orthonormal vectors, which we may call $\mathbf{u}_0$ and $\mathbf{u}_1$ in what follows, in $\mathbb{C}^{4}$. Before we proceed any further, let us count the total number of available degrees of freedom. On the one hand, we have four real numbers whose squares add up to one for $\mathbf{u}_0$, which amounts to 3 degrees of freedom. For $\mathbf{u}_1$, we have four complex numbers with three constraints (one for normalisation and two for the orthogonality with $\mathbf{u}_0$), which results in 5 degrees of freedom. In total, we need 8 parameters per system qubit.

Our parametrisation for single-qubit POVMs thus consists of 8 real numbers $\vec{x} = (x_0, \ldots, x_7)$, with $x_i \in (0, 1), \, \forall i$ (in practice, we constrain the values further, see App.~\ref{app:gradient_descent}). We start by using the first three of these to produce the set of angles $(\pi x_0, \pi x_1, 2 \pi x_2)$, which identify (uniquely) a point on a 3-sphere $\mathbb{S}^{3}$ with unit radius embedded in $\mathbb{R}^{4}$. The corresponding Euclidean coordinates in the embedding space are four real numbers whose squares add up to one, hence generating $u_{00}^{b_q b_a}$. Defining $u_{10}^{b_q b_a}$ from the other five parameters is slightly more involved. To guarantee that the vector $\mathbf{u}_1$ is orthogonal to $\mathbf{u}_0$, we construct it as a linear combination of orthonormal vectors orthogonal to $\mathbf{u}_0$, that is, $\mathbf{u}_1 = \sum_i z_i \mathbf{u}^{\perp}_i$; the orthonormal basis $\lbrace \mathbf{u}^{\perp}_i \rbrace$ can be found by means of the Gram-Schmidt orthonormalisation. The components $z_i$, which must also be normalised, are determined by the remaining parameters: once again, we define a list of angles $(\pi x_3, \ldots, \pi x_6, 2 \pi x_7)$ and calculate the Euclidean coordinates of the corresponding point in $\mathbb{S}^{5}$. These six real numbers $\lbrace r_i, \, i \in [0, 5] \rbrace$ are then used to define three components $\lbrace z_k = r_{2 k} + i r_{2 k + 1} \rbrace$. The result of this procedure is a vector $\mathbf{u}_1 \in \mathbb{C}^{4}$ whose components can be identified with $u_{10}^{b_q b_a}$.

Finally, we must find two more vectors $\mathbf{u}_2, \mathbf{u}_3 \in \mathbb{C}^{4}$ to complete the missing terms $u_{k1}^{b_q b_a}$ in the definition of the unitary. This can be done by using the Gram-Schmidt orthonormalisation once more. Once the unitary $U$ is defined, we can not only calculate the corresponding set of effects $\lbrace \Pi_i \rbrace$, but also implement it in a given circuit. Indeed, the algorithms to find the circuit decomposition of unitary $U$ are known and readily implemented in Qiskit~\cite{qiskit} (also, note that any two-qubit gate can be decomposed in up to three CNOT gates).

Admittedly, this methodology is more complicated than simply parametrising arbitrary two-qubit gates $U$ and then calculating the corresponding POVM. However, as discussed above, our procedure avoids the use of unnecessary or redundant parameters, which could make the POVM optimization harder. Nevertheless, it is likely that other parametrisations, more suitable for the adaptive optimization algorithm, exist. These refinements, as well as improving the gradient descent protocol (see App.~\ref{app:gradient_descent}), will be the subject of future work.

To conclude this section, let us also explain the geometric representation of the POVMs in Fig.~\ref{fig:energy_estimation}. Note that the effects introduced above are rank-1, so they can be written as $\Pi_i = \gamma \tilde{\Pi}_i$, where $\gamma = \mathrm{Tr} [\Pi_i]$ and $\tilde{\Pi}_i$ are single-qubit pure states (also, $\gamma \leq 1$). As a result, we can exploit the Bloch sphere representation and write $\Pi_i (\vec{r}) = \gamma (\mathbb{I} + \vec{q} \cdot \vec{\sigma}) / 2$ with $\vec{q} = (\mathrm{Tr} [\tilde{\Pi}_i \sigma_x], \mathrm{Tr} [\tilde{\Pi}_i \sigma_y], \mathrm{Tr} [\tilde{\Pi}_i \sigma_z])$ and $\vert \vec{q} \vert = 1$. Thus, defining $\vec{r} = (\mathrm{Tr} [\Pi_i \sigma_x], \mathrm{Tr} [\Pi_i \sigma_y], \mathrm{Tr} [\Pi_i \sigma_z]) = \gamma \vec{q}$, we have $\Pi_i (\vec{r}) = (\vert \vec{r} \vert \mathbb{I} + \vec{r} \cdot \vec{\sigma}) / 2$.

\section{Gradient descent protocol}\label{app:gradient_descent}

Along the measurement process, we iteratively update the POVM parameters as well as the number of shots per experiment. In particular, we gradually increase the number of shots  in order to have more precise estimations of the second moment as the POVM parameters approach a minimum and, consequently, the gradient decreases in magnitude. In this section, we briefly outline the protocol used in our numerical experiments.

As explained in the main text, the POVM-based measurements allow us to estimate the gradient $\nabla_{\vec{x}} \langle \omega^2_\mathbf{m} \rangle$ classically from the outcomes of an experiment run with the POVM corresponding to parameters $\vec{x}_t$, where $t$ labels the iteration (for the finite-difference partial derivatives $\partial_{x_k} \langle \omega_\mathbf{m}^2 \rangle \approx (\langle {\omega'}^2_\mathbf{r} \rangle - \langle \omega_\mathbf{m}^2 \rangle) / h$, we use $h = 10^{-3}$). With these elements, we determine the POVM to be used in the $(t+1)$-th iteration through $\vec{x}_{t+1} = \vec{x}_t - \nu \nabla_{\vec{x}} \langle \omega^2_\mathbf{m} \rangle / \max (\vert \nabla_{\vec{x}} \langle \omega^2_\mathbf{m} \rangle \vert)$, where $\vert \nabla_{\vec{x}} \langle \omega^2_\mathbf{m} \rangle \vert$ is to be understood as the set of absolute values of the components of $\nabla_{\vec{x}} \langle \omega^2_\mathbf{m} \rangle$. Hence, $\nu$ is the magnitude of the largest change, in absolute value, of the POVM parameters. It should also be mentioned that, to avoid numerical instabilities, we further constrain every parameter to be between $[\delta, 1 - \delta]$, with $\delta = 0.05$. We start our simulations with $S_1 = 1000$ shots, and we use $\nu = 0.05$. Every three iterations, we update $S_t + 1000 \rightarrow S_{t+1}$ and $\nu / 1.2 \rightarrow \nu$. Hence, as the algorithm approaches the minimum, we obtain more precise estimations of the gradient (larger $S_t$) and we make smaller changes to the parameters (smaller $\nu$).

This parameter updating schedule is rather heuristic and still leaves room for improvement. Designing a more theory-driven approach, or using more sophisticated optimization techniques, will be the subject of future work.

\section{Symmetric IC POVMs as initial measurements and correlated estimators}\label{app:sic-povms}

In the absence of prior knowledge about the state of the qubit register, it is desirable to use a so-called symmetric informationally complete POVM (SIC POVM) on every system qubit. Symmetric here means that its single-qubit effects, when rescaled as $\tilde{\Pi}_i = 2 \Pi_i$ yield a set of projectors $\lbrace \tilde{\Pi}_i : \tilde{\Pi}^2_i = \tilde{\Pi}_i \rbrace$ fulfilling $\mathrm{Tr}[\tilde{\Pi}_i \tilde{\Pi}_j] = (2 \delta_{ij} + 1) / 3), \, \forall i, j$. Hence, the projectors $\lbrace \tilde{\Pi}_i \rbrace$ form a regular tetrahedron in the Bloch sphere.

In this work, we have considered two different SIC POVMs as initial conditions for the adaptive algorithm. The first one is the classic example of single-qubit SIC POVM, defined in terms of the projectors $\lbrace \tilde{\Pi}_i = \ket{\tilde{\pi}_i} \bra{\tilde{\pi}_i} : \ket{\tilde{\pi}_0} = \ket{0}, \, \ket{\tilde{\pi}_k} = (\ket{0} + \sqrt{2} e^{i 2 \pi (k-1) / 3} \ket{1}) / \sqrt{3}, \, k \in [1, 3] \rbrace$. The second SIC POVM used in this paper was considered by Jiang et al.~\cite{Jiang2020optimalfermionto} and is another standard setting \cite{Renes2004,Rehacek2004, axioms6030021}. In order to use them in our algorithm, we must first find the parameters $\vec{x}$ of each of them in the POVM space (see App.~\ref{app:povm_implementation}). This can be done numerically; the resulting parameters are reported in the computer code accompanying this paper \cite{sourcecode}.

It is worth discussing some properties of this second SIC POVM when used in our hybrid quantum-classical Monte Carlo algorithm, Eq.~\eqref{eq:estimated_mean}. In this case, all the $b^{(i)}_{km}$ exhibit the nice feature $b^{(i)}_{km} \in \lbrace -\sqrt{3}, \sqrt{3} \rbrace, \, \forall k > 0$ (for $k = 0$, these are equal to one, since the effects add up to identity). This, in turn, has interesting implications. Let us consider the statistical error in the estimation of the expectation value of an observable given by a single Pauli string $P_\mathbf{k}$ with weight $l$, that is, only $l$ Pauli operators in $P_\mathbf{k}$ are different from identity. In this case, the variance of the Monte Carlo is given by $\mathrm{Var} (\omega_\mathbf{m}) = 3^l - \langle P_\mathbf{k} \rangle^2 \leq 3^l$. Hence, if $S$ measurements are performed, the variance of the estimator $\bar{P}_\mathbf{k}$ is $\mathrm{Var} (\bar{P}_\mathbf{k}) \leq 3^l / S$. This is indeed consistent with Ref.~\cite{Jiang2020optimalfermionto}.

\begin{figure*}[t]
\begin{minipage}[!b]{0.65\textwidth}
    \includegraphics[width=\textwidth]{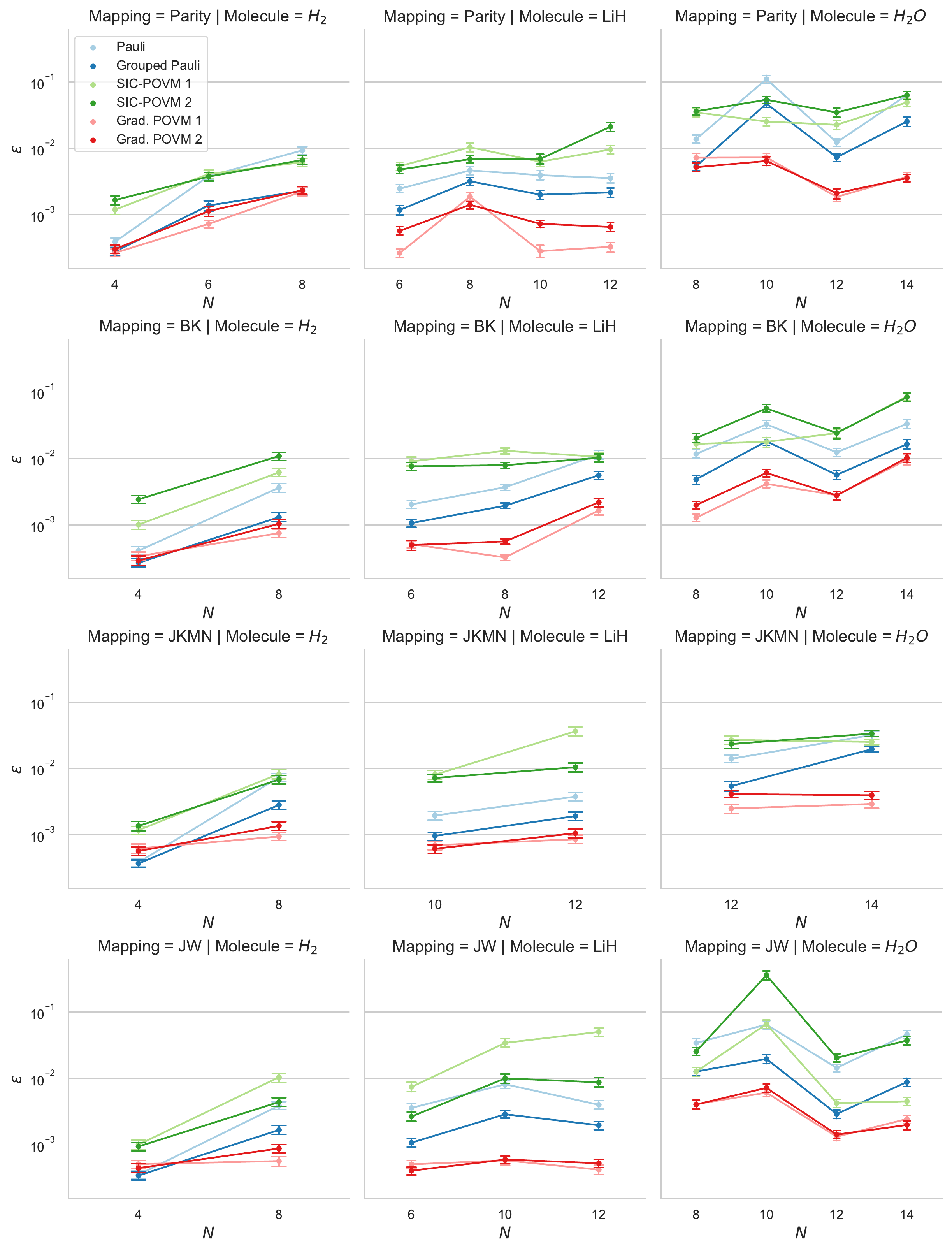}
\end{minipage}
\begin{minipage}[b]{0.3\textwidth}
\resizebox{5cm}{!}{
\begin{tabular}{llrlccc}
\toprule
Mol.       & Mapping   &  $N$    &  Basis &  TQR &  $Z_2$ & CF \\
\hline
H2 & BK &       4 &  STO3G &                 &                  &         \\
    & BK &       8 &   631G &                 &                  &         \\
    & JW &       4 &  STO3G &                 &                  &         \\
    & JW &       8 &   631G &                 &                  &         \\
    & JKMN &       4 &  STO3G &                 &                  &         \\
    & JKMN &       8 &   631G &                 &                  &         \\
    & Parity &       4 &  STO3G &                 &                  &         \\
    & Parity &       6 &   631G &                 \checkmark &                  &         \\
    & Parity &       8 &   631G &                 &                  &         \\
    \hline
LiH & BK &       6 &  STO3G &                 &                  \checkmark &         \checkmark \\
    & BK &       8 &  STO3G &                 &                  \checkmark &         \\
    & BK &      12 &  STO3G &                 &                  &         \\
    & JW &      12 &  STO3G &                 &                  &         \\
    & JW &       6 &  STO3G &                 &                  \checkmark &         \checkmark \\
    & JW &      10 &  STO3G &                 &                  &         \checkmark \\
    & JKMN &      12 &  STO3G &                 &                  &         \\
    & JKMN &      10 &  STO3G &                 &                  &         \checkmark \\
    & Parity &       8 &  STO3G &                 \checkmark &                  \checkmark &         \\
    & Parity &       6 &  STO3G &                 \checkmark &                  \checkmark &         \checkmark \\
    & Parity &      10 &  STO3G &                 \checkmark &                  &         \\
    & Parity &      12 &  STO3G &                 &                  &         \\
    \hline
H2O & BK &      10 &  STO3G &                 &                  \checkmark &         \\
    & BK &       8 &  STO3G &                 &                  \checkmark &         \checkmark \\
    & BK &      12 &  STO3G &                 &                  &         \checkmark \\
    & BK &      14 &  STO3G &                 &                  &         \\
    & JW &      12 &  STO3G &                 &                  &         \checkmark \\
    & JW &      14 &  STO3G &                 &                  &         \\
    & JW &       8 &  STO3G &                 &                  \checkmark &         \checkmark \\
    & JW &      10 &  STO3G &                 &                  \checkmark &         \\
    & JKMN &      12 &  STO3G &                 &                  &         \checkmark \\
    & JKMN &      14 &  STO3G &                 &                  &         \\
    & Parity &       8 &  STO3G &                 &                  \checkmark &         \checkmark \\
    & Parity &      10 &  STO3G &                 &                  \checkmark &         \\
    & Parity &      12 &  STO3G &                 &                  &         \checkmark \\
    & Parity &      14 &  STO3G &                 &                  &         \\
\botrule
\end{tabular}
}
\end{minipage}\hfill
    \caption{(Left) Error in the estimation of the energy of an optimal VQE circuit for all the Hamiltonians reported in the table on the right. Every column compares the results for one molecule ($\rm{H_2}$, $\rm{LiH}$, and $\rm{H_2O}$) with different measurement methods, with a total of $S = 10^6$ shots. Each row corresponds to a different mapping. Points represent the average error over 100 realisations and the error bars show a $95\%$ confidence interval obtained using bootstrapping. We note that, in some cases, especially those involving more qubits, like the 14-qubit $\rm{H_2O}$ molecule with the BK mapping, the measurement optimization has not fully converged for $S = 10^{6}$, so one would expect more notable differences with respect to the other methods for larger $S$. (Right) A table of the various combinations of molecule, mapping, basis and qubit-reduction techniques considered, with the corresponding number of qubits $N$. TQR is the two-qubit reduction for the parity mapping, $Z_2$ refers to qubit reductions due to discrete symmetries \cite{Bravyi17} and CF denotes core freeze.}
    \label{fig:all_results}
\end{figure*}

While we can reuse the IC data from the quantum computer to calculate the expectation value of other Pauli strings $P_\mathbf{k'}$ with similar statistical error (assuming they have the same weight $l$), we must take into account that the resulting estimators $\bar{P}_\mathbf{k}$ and $\bar{P}_\mathbf{k'}$ can be correlated. In practice, this means that, if we are to use them to calculate the expectation value of an operator defined in terms of a linear combination of Pauli strings,  $\mathcal{O} = \sum_\mathbf{k} c_\mathbf{k} P_\mathbf{k}$, the variance of the estimator  $\bar{\mathcal{O}} = \sum_\mathbf{k} c_\mathbf{k} \bar{P}_\mathbf{k}$ depends on the potentially non-zero covariance between distinct terms, so we cannot assume that $\mathrm{Var} (\bar{\mathcal{O}}) = \sum_\mathbf{k} \vert c_\mathbf{k} \vert^2 \mathrm{Var} (\bar{P}_\mathbf{m})$.

The estimation based on the Monte Carlo method, Eq.~\eqref{eq:estimated_mean}, naturally takes into account these correlations when accounting for the statistical error of the approach, hence yielding the correct estimation. This is important for two reasons. On the one hand, it provides a meaningful assessment of how far the algorithm is from reaching the required accuracy at any given point of its execution. On the other hand, since the Monte Carlo variance is the quantity that our adaptive strategy seeks to minimize, the algorithm presented here can potentially find POVMs for which the negative impact of these correlations on the estimated mean is reduced.

\section{Sequential and one-step mixing equivalence}\label{app:sequential_vs_one-step}

In this section we prove that the sequential estimation mixing presented in the main text is unbiased. To show this, let us first compute the unbiased one-step mixing estimation. Suppose that, after the different experiments have been run, we are left with a set of $T$ estimated means $\lbrace \bar{\mathcal{O}}_t \rbrace$ and variances $\lbrace \bar{V}_t \rbrace$. We would like to find a set of weights $\lbrace \alpha_t > 0 \rbrace$, with $\sum_t \alpha_t = 1$, that minimizes the variance $\bar{\bar{V}}_T = \sum_t \alpha_t^2 \bar{V}_t$ of $\bar{\bar{\mathcal{O}}}_T = \sum_t \alpha_t \bar{\mathcal{O}}_t$. To do so, we can introduce a Lagrange multiplier $\lambda$ and define
\begin{equation}
\mathcal{L} = \sum \limits_{t=1}^{T} \alpha_t^2 \bar{V}_t - \lambda \left( \sum_t \alpha_t - 1\right),
\end{equation}
so that $\partial_\lambda \mathcal{L} = 0$ imposes the constraint $\sum_t \alpha_t = 1$. From $\partial_{\alpha_t} \mathcal{L} = 0$ we obtain $\alpha_t = \lambda \bar{\rho}_t / 2$, where we have defined $\bar{\rho}_t \equiv 1 / \bar{V}_t$ to ease the presentation, as inverse variances will appear throughout. Using now $\sum_t \alpha_t = 1$ yields $\lambda = 2 / \sum_i \bar{\rho}_i$ and $\alpha_t = \bar{\rho}_t / \sum_i \bar{\rho}_i$. Hence, we arrive at
\begin{equation}\label{eq:one_step_mixing}
\bar{\bar{\mathcal{O}}}_T = \frac{\sum \limits_{t=1}^{T} \bar{\mathcal{O}}_t \bar{\rho}_t}{\sum \limits_{t=1}^{T} \bar{\rho}_t} \quad \mathrm{and} \quad \bar{\bar{V}}_T = \frac{1}{\sum \limits_{t=1}^{T} \bar{\rho}_t}.
\end{equation}

To assess the result of the sequential algorithm, note that the recurrence $\bar{\bar{V}}_{t} = \bar{\bar{V}}_{t-1} \bar{V}_{t} / (\bar{\bar{V}}_{t-1} + \bar{V}_{t})$ in the second step is equivalent to $\bar{\bar{\rho}}_t = \bar{\bar{\rho}}_{t-1} + \bar{\rho}_t$, with $\bar{\bar{\rho}}_t \equiv 1 / \bar{\bar{V}}_t$. Iterating, we obtain $\bar{\bar{\rho}}_T = \sum_{t=1}^T \bar{\rho}_t$, which is the right-most term in Eq.~\eqref{eq:one_step_mixing}. Similarly, the recurrence for the mean, $\bar{\bar{\mathcal{O}}}_t = (\bar{\mathcal{O}}_{t} \bar{\bar{V}}_{t-1} + \bar{\bar{\mathcal{O}}}_{t-1} \bar{V}_{t}) / (\bar{\bar{V}}_{t-1} + \bar{V}_{t})$, reads $
\bar{\bar{\mathcal{O}}}_t = \left[ \bar{\mathcal{O}}_{t} \bar{\rho}_t + \bar{\bar{\mathcal{O}}}_{t-1} \bar{\bar{\rho}}_{t-1} \right] / \bar{\bar{\rho}}_t$. Iterating once again, we obtain the expression for $\bar{\bar{\mathcal{O}}}_T$ in Eq.~\eqref{eq:one_step_mixing}. Hence, both estimations are equivalent.

\section{Other experiments}\label{app:other_experiments}
In the main text we presented a selection of the numerical results obtained in this work allowing us to showcase the main features of our method. For completeness, in this Appendix we report all the results of the simulations obtained with the $\textrm{H}_2$, $\textrm{LiH}$ and $\mathrm{H_2O}$ molecules, using different combinations of fermion-to-qubit mappings (namely parity, JW, BK and JKMN) and qubit reduction techniques (namely two-qubit parity reduction, $Z_2$ symmetry and core freeze), where applicable. Figure~\ref{fig:all_results} shows the average absolute error $\varepsilon$ on the estimation of the energy with $10^6$ shots, for the Hamiltonians reported in the table. The plots are in line with the figures in the main text, as our algorithm generally reaches smaller errors with the same total number of measurements. All the data used in this manuscript are available on Zenodo \cite{dataset}. The source code used to generate the molecular Hamiltonians and all the results presented here is available online \cite{sourcecode}.

\bibliography{bib_meas.bib}

\end{document}